\RequirePackage{tablefootnote}
\documentclass[twocolumn,linenumbers]{aastex631}

\usepackage{threeparttable}
\usepackage{multirow}

\begin{document}

\title{Strict limits on potential secondary atmospheres on the temperate rocky exo-Earth TRAPPIST-1\,d}
\nolinenumbers

\correspondingauthor{Caroline Piaulet-Ghorayeb}
\email{carolinepiaulet@uchicago.edu}

\author[0000-0002-2875-917X]{Caroline Piaulet-Ghorayeb}
\altaffiliation{E. Margaret Burbridge Postdoctoral Fellow}
\affiliation{Trottier Institute for Research on Exoplanets and Department of Physics, Universit\'{e} de Montr\'{e}al, Montreal, QC, Canada}
\affiliation{Department of Astronomy \& Astrophysics, University of Chicago, 5640 South Ellis Avenue, Chicago, IL 60637, USA}

\author[0000-0001-5578-1498]{Bj\"{o}rn Benneke}
\affiliation{Trottier Institute for Research on Exoplanets and Department of Physics, Universit\'{e} de Montr\'{e}al, Montreal, QC, Canada}
\affiliation{Department of Earth, Planetary, and Space Sciences, University of California, Los Angeles, CA 90095, USA}

\author[0000-0003-2260-9856]{Martin Turbet}
\affiliation{Laboratoire de M\'et\'eorologie Dynamique/IPSL, CNRS, Sorbonne Universit\'e, Ecole Normale Sup\'erieure, PSL Research University, Ecole Polytechnique, 75005 Paris, France}
\affiliation{Laboratoire d'astrophysique de Bordeaux, Univ. Bordeaux, CNRS, B18N, allée Geoffroy Saint-Hilaire, 33615 Pessac, France}

\author{Keavin Moore}
\affiliation{Department of Physics, McGill University, Montr\'{e}al, QC, Canada}
\affiliation{Department of Earth and Planetary Sciences, McGill University, Montr\'{e}al, QC, Canada}

\author[0000-0001-6809-3520]{Pierre-Alexis Roy}
\affiliation{Trottier Institute for Research on Exoplanets and Department of Physics, Universit\'{e} de Montr\'{e}al, Montreal, QC, Canada}

\author[0000-0003-4676-0622]{Olivia Lim}
\affiliation{Trottier Institute for Research on Exoplanets and Department of Physics, Universit\'{e} de Montr\'{e}al, Montreal, QC, Canada}

\author[0000-0001-5485-4675]{Ren\'{e} Doyon}
\affiliation{Trottier Institute for Research on Exoplanets and Department of Physics, Universit\'{e} de Montr\'{e}al, Montreal, QC, Canada}

\author[0000-0002-5967-9631]{Thomas J. Fauchez} 
\affiliation{NASA Goddard Space Flight Center,
8800 Greenbelt Road,
Greenbelt, MD 20771, USA}
\affiliation{Integrated Space Science and Technology Institute, Department of Physics, American University, Washington DC}
\affiliation{NASA GSFC Sellers Exoplanet Environments Collaboration}

\author[0000-0003-0475-9375]{Lo\"{i}c Albert}
\affiliation{Trottier Institute for Research on Exoplanets and Department of Physics, Universit\'{e} de Montr\'{e}al, Montreal, QC, Canada}

\author[0000-0002-3328-1203]{Michael Radica} 
\altaffiliation{NSERC Postdoctoral Fellow}
\affiliation{Trottier Institute for Research on Exoplanets and Department of Physics, Universit\'{e} de Montr\'{e}al, Montreal, QC, Canada}
\affiliation{Department of Astronomy \& Astrophysics, University of Chicago, 5640 South Ellis Avenue, Chicago, IL 60637, USA}

\author[0000-0002-2195-735X]{Louis-Philippe Coulombe}
\affiliation{Trottier Institute for Research on Exoplanets and Department of Physics, Universit\'{e} de Montr\'{e}al, Montreal, QC, Canada}

\author[0000-0002-6780-4252]{David Lafreni\`{e}re}
\affiliation{Trottier Institute for Research on Exoplanets and Department of Physics, Universit\'{e} de Montr\'{e}al, Montreal, QC, Canada}

\author[0000-0001-6129-5699]{Nicolas B. Cowan}
\affiliation{Department of Physics, McGill University, Montr\'{e}al, QC, Canada}
\affiliation{Department of Earth and Planetary Sciences, McGill University, Montr\'{e}al, QC, Canada}

\author[0009-0007-6298-9802]{Danika Belzile}
\affiliation{Department of Physics, Universit\'{e} de Montr\'{e}al, Montr\'{e}al, QC, Canada}
\affiliation{Dawson College, Montr\'{e}al, QC, Canada}

\author[0009-0007-6653-6165]{Kamrul Musfirat}
\affiliation{Department of Physiology, McGill University, Montr\'{e}al, QC, Canada}

\author[0009-0004-9827-2481]{Mehramat Kaur}
\affiliation{École secondaire Cavalier-De LaSalle, Montr\'{e}al, QC, Canada}

\author[0009-0005-6135-6769]{Alexandrine L'Heureux}
\affiliation{Trottier Institute for Research on Exoplanets and Department of Physics, Universit\'{e} de Montr\'{e}al, Montreal, QC, Canada}

\author[0000-0002-6773-459X]{Doug Johnstone}
\affiliation{NRC Herzberg Astronomy and Astrophysics, 5071 West Saanich Rd,
Victoria, V9E 2E7, BC, Canada}
\affiliation{Department of Physics \& Astronomy, University of Victoria, Victoria,
V8P 5C2, BC, Canada}

\author[0000-0003-4816-3469]{Ryan J. MacDonald}
\altaffiliation{NHFP Sagan Fellow}
\affiliation{Department of Astronomy, University of Michigan, 1085 S. University Ave., Ann Arbor, MI 48109, USA}

\author[0000-0002-1199-9759]{Romain Allart}
\affiliation{Trottier Institute for Research on Exoplanets and Department of Physics, Universit\'{e} de Montr\'{e}al, Montreal, QC, Canada}
\thanks{SNSF Postdoctoral Fellow}

\author[0000-0003-4987-6591]{Lisa Dang}
\affiliation{Trottier Institute for Research on Exoplanets and Department of Physics, Universit\'{e} de Montr\'{e}al, Montreal, QC, Canada}



\author[0000-0002-0436-1802]{Lisa Kaltenegger}
\affiliation{Carl Sagan Institute, Cornell University, Ithaca, NY, USA}
\affiliation{Cornell Center for Astrophysics and Planetary Science, Cornell University, Ithaca, NY 14853, USA}
\affiliation{Astronomy Department, Cornell University, Ithaca, NY 14853, USA}

\author[0000-0002-8573-805X]{Stefan Pelletier}
\affiliation{Observatoire astronomique de l’Université de Genève, 51 chemin Pegasi 1290 Versoix, Switzerland}
\affiliation{Trottier Institute for Research on Exoplanets and Department of Physics, Universit\'{e} de Montr\'{e}al, Montreal, QC, Canada}

\author[0000-0002-5904-1865]{Jason F. Rowe}
\affiliation{Department of Physics and Astronomy, Bishop's University, 2600 Rue College, Sherbrooke, QC J1M 1Z7, Canada}

\author[0000-0003-4844-9838]{Jake Taylor}
\affiliation{Department of Physics, University of Oxford, Parks Rd, Oxford OX1 3PU, UK}
\affiliation{Trottier Institute for Research on Exoplanets and Department of Physics, Universit\'{e} de Montr\'{e}al, Montreal, QC, Canada}

\author[0000-0001-7836-1787]{Jake D. Turner}
\affiliation{Carl Sagan Institute, Cornell University, Ithaca, NY, USA}
\affiliation{Astronomy Department, Cornell University, Ithaca, NY 14853, USA}




\begin{abstract}
\begin{nolinenumbers}

The nearby TRAPPIST-1 system, with its seven small rocky planets orbiting a late-type M8 star, offers an unprecedented opportunity to search for secondary atmospheres on temperate terrestrial worlds. In particular, the 0.8 R$_\oplus$ TRAPPIST-1\,d lies at the edge of the habitable zone ($T_\mathrm{eq,A=0.3} = 262$ K). Here we present the first 0.6–5.2 $\mu$m NIRSpec/PRISM transmission spectrum of TRAPPIST-1\,d from two transits with \textit{JWST}. We find that stellar contamination from unocculted bright heterogeneities introduces 500–1,000 ppm visit-dependent slopes, consistent with constraints from the out-of-transit stellar spectrum. Once corrected, the transmission spectrum is flat within $\pm 100$–150 ppm, showing no evidence for a haze-like slope or molecular absorption despite NIRSpec/PRISM's sensitivity to CH$_4$, H$_2$O, CO, SO$_2$, and CO$_2$. Our observations exclude clear, hydrogen-dominated atmospheres with high confidence ($>3\sigma$). We leverage our constraints on even trace amounts of CH$_4$, H$_2$O, and CO$_2$ to further reject high mean molecular weight compositions analogous to Titan, a cloud-free Venus, early Mars, and both Archean Earth and a cloud-free modern Earth scenario ($>$95\% confidence). If TRAPPIST-1\,d retains an atmosphere, it is likely extremely thin or contains high-altitude aerosols, with water cloud formation at the terminator predicted by 3D global climate models. Alternatively, if TRAPPIST-1\,d is airless, our evolutionary models indicate that TRAPPIST-1 b, c, and d must have formed with $\lesssim$4 Earth oceans of water, though this would not preclude atmospheres on the cooler habitable-zone planets TRAPPIST-1 e, f, and g.
\end{nolinenumbers}

\end{abstract}




\section{Introduction} \label{sec:intro}

TRAPPIST-1 is a nearby (12 pc) small (0.12 $R_\odot$, see Table \ref{table:pla_star_param}; \citealp{agol_refining_2021}) ultra-cool M8V star \citep{liebert_ri_2006, gillon_seven_2017,davoudi_updated_2024}, host to seven transiting rocky planets receiving moderate irradiation \citep{gillon_seven_2017}. The extremely small size of the star boosts the detectability of even high mean molecular weight secondary atmospheres \citep{morley_forward_2017,lincowski_evolved_2018,lustig-yaeger_detectability_2019} on top of the small TRAPPIST-1 planets (approximately 0.75 to 1.1 $R_\oplus$), and shrinks the orbital distance where water can be in the liquid form (``habitable zone'') relative to Sun-like stars. This offers frequent opportunities to probe the atmospheres of potentially habitable planets on orbital periods of only about 4 to 10 days. 

TRAPPIST-1\,d is the third planet from the host star. At an orbital distance of a mere 0.02 AU, the exo-Earth (0.8 R$_\oplus$, see Table \ref{table:pla_star_param}) receives only 1.12 times as much flux from its host star as Earth receives from the Sun \citep{gillon_seven_2017}. Because of its proximity to its host star, however, TRAPPIST-1\,d is believed to be tidally locked with a permanent night- and dayside \citep{turbet_review_2020}. 
TRAPPIST-1\,d is predicted by 1D models to lie inside the runaway greenhouse limit \citep{kopparapu_habitable_2013,gillon_seven_2017,Massol_2023}. However, 3D global climate models (GCMs) predict that the radiative feedback and reflectivity of dayside clouds may allow for liquid water conditions at its surface given its synchronously rotating state \citep{yang_stabilizing_2013,kopparapu_inner_2016}, if cloud properties are favorable \citep{wolf_assessing_2017,turbet_modeling_2018,way_trappist-1_2025}. The presence of surface liquid water, however, presupposes that TRAPPIST-1\,d cooled down enough after its formation to allow for water to condense out of the atmosphere, which is unlikely at its present-day orbital distance \citep{turbet_water_2023}. 
With its short (4 days) orbital period which enables many repeat observations, and its low mass (0.39 Earth masses, compared to 0.69 and 1.04 Earth masses for the outer planets e and f), TRAPPIST-1\,d is an extremely favorable small temperate planet for transmission spectroscopy \citep{hill_catalog_2023}. Overall, this planet stands out as a great candidate for an observational test of the habitable zone concept, as different model assumptions lead to a range of predictions about its composition and surface conditions. 


The seven planets' near-resonant chain produces transit-timing variations (TTVs; \citealt{grimm_nature_2018,agol_refining_2021,wang_updated_2017}) that enable precise mass measurements, which would be challenging to obtain using radial velocities. The latest determination of planet masses and radii in the TRAPPIST-1 system demonstrated that, within measurement uncertainties, all seven planets can be explained by the same iso-composition curve \citep{agol_refining_2021}. They, however, have consistently lower bulk densities than the solar system terrestrial planets, indicating either lower amounts of iron in their interior or moderate enrichment in volatile species such as water. The TRAPPIST-1 planets could even be ``core-free'' planets \citep{elkins-tanton_ranges_2008}, where all the iron has oxidized and remains within the mantle rather than segregated in a metal core. While the planets' bulk densities are similar, they show a tentative trend of lower density with increasing stellocentric distance, suggesting a gradient in volatile endowment. In this trend, TRAPPIST-1\,d stands out as an outlier, with a lower density than expected for a composition consistent with planets c and e. This leads to varying estimates of the water mass fraction of planet d, from less than 0.01\% assuming water is in the vapor form \citep{agol_refining_2021}, and
up to a few percent assuming the surface conditions allow for liquid water to form \citep{acuna_characterisation_2021}. 
Such volatiles could have been provided by a formation near the ice line, through pebble accretion \citep{unterborn_inward_2018,coleman_pebbles_2019}, or if it formed in a ``cold finger'' of the protoplanetary disk where water ice was locally more abundant \citep{cyr_distribution_1998}.


Recent explorations with JWST of the two innermost TRAPPIST-1 planets, TRAPPIST-1\,b \citep{ih_constraining_2023,lim_atmospheric_2023,greene_thermal_2023,ducrot_combined_2024} and TRAPPIST-1\,c \citep{zieba_no_2023,lincowski_potential_2023,RadicaT1c}, revealed poor heat redistribution (in emission) and flat spectra (in transmission), to the level of excluding some high mean molecular weight scenarios, overall favoring a bare-rock interpretation. Another key finding from atmosphere searches on the TRAPPIST-1 planets (see e.g., \citealt{de_wit_atmospheric_2018,zhang_near-infrared_2018,wakeford_disentangling_2019,garcia_hstwfc3_2022,lim_atmospheric_2023,RadicaT1c,rathcke_stellar_2025}) is that potential planetary molecular absorption features in the measured transmission spectra are overwhelmed by 200--300 ppm amplitude signals that are varying from one visit to the next, and are likely caused by the transit light source (TLS) effect. The TLS effect is caused by the presence of patches of the unocculted stellar surface (e.g. spots, faculae) that are cooler or hotter than the photosphere along the planet's transit path \citep{sing_hubble_2011,berta_gj1214_2011,rackham_transit_2018}. Although this effect can be marginalized over (see e.g. \citealt{fournier-tondreau_near-infrared_2023}), it requires accurate models for the stellar spectrum and heterogeneity components, which are not currently available for M dwarfs in general although 3D models are becoming available for early-type G, K, and M dwarfs (e.g. \citealp{smitha_first_2025}). For the late-type M8 star TRAPPIST-1 in particular, the stellar spectrum exhibits features that are only partially matched by models \citep{lim_atmospheric_2023, RadicaT1c}.
The M8-type stellar host also poses other challenges from an observational standpoint, including frequent stellar flares, with variable shapes and wavelength-dependent signatures, that often affect space-based observations \citep{howard_characterizing_2023}.

In terms of atmosphere prospects, contrary to TRAPPIST-1\,d, planets b and c were already a priori expected to be stripped of an atmosphere through hydrodynamic escape because they lie well within the runaway greenhouse limit, receiving 4.2 and 2.2 times the Earth's insolation flux \citep{agol_refining_2021,krissansen-totton_predictions_2022}. Even assuming that TRAPPIST-1\,b and c are airless, accounting for the shared irradiation history, magma ocean geochemical and thermal evolution, and plausible initial volatile budgets of the seven planets leads to about 70\% probability of atmospheric retention for TRAPPIST-1\,d \citep{krissansen-totton_implications_2023,gialluca_implications_2024} - although other models are more pessimistic \citep{van_looveren_airy_2024}. Whether the planet retains an atmosphere or not, however, depends on its initial volatile inventory. Atmospheric reconnaissance can therefore serve as a powerful constraint on its initial formation conditions, and is also motivated by the growing community effort \citep{redfield_DDT_2024} to constrain the location of the ``cosmic shoreline'' \citep{zahnle_cosmic_2017}: the region of parameter space where rocky planets can retain or revive atmospheres. 

Due to the larger orbital distance and cooler equilibrium temperature of TRAPPIST-1\,d compared to planets b and c, transmission spectroscopy is the method of choice for atmospheric reconnaissance of this planet. TRAPPIST-1\,d has already been the target of multiple transmission spectroscopy campaigns with the Hubble and Spitzer space telescopes \citep{de_wit_atmospheric_2018,ducrot_0.8-4.5mum_2018,ducrot_trappist-1_2020}. So far, observations have only ruled out cloud-free, solar-metallicity low mean molecular weight atmospheres, while not reaching the sensitivity required to search for more compact, high mean molecular weight atmospheres. 
With its high precision and wide wavelength coverage, the PRISM mode of JWST's near-infrared spectrograph (NIRSpec) instrument \citep{Birkmann_NIRSPEC_2022} has the potential to detect and identify potential high mean molecular weight atmospheres, as well as tell apart the wavelength-dependent signatures of clouds or hazes from those of molecular species. At the longest wavelengths, the TLS effect is also expected to play a minor role in contaminating the planetary spectrum \citep{seager_why_2024}.

Here we present the first JWST transmission spectrum of TRAPPIST-1\,d, obtained with NIRSpec/PRISM from two transit observations. We first outline the observations and data reduction in Section \ref{sec:obs}, and describe the light-curve fitting in Section \ref{sec:lc_analysis}. Our methods for modeling the contributions of different stellar surface components to the transmission spectra and the out-of-transit stellar spectra are described in Section \ref{sec:stellar_ctm} and discussed in Section \ref{sec:complementarity_oot_TLS}. We present our methods and results from atmospheric retrievals in Section \ref{sec:atm_modeling}. We discuss our results in Section \ref{sec:disc} and conclude in Section \ref{sec:conclusion}.

\section{Observations and Data Reduction}\label{sec:obs}


\subsection{Summary of the observations}\label{sec:nirspec_obs}

JWST observed two consecutive transits of TRAPPIST-1\,d with NIRSpec/PRISM on UTC November 5 and November 9, 2022. Both transits were observed as part of the Cycle 1 ``NIRISS Exploration of the Atmospheric diversity of Transiting exoplanets (NEAT) Guaranteed Time Observations program GTO 1201 (PI: David Lafreni\`{e}re). Each NIRSpec/PRISM Bright Object Time Series (BOTS) transit observation lasted 3.2 hours and consisted of 7,200 integrations with 6 groups each, i.e., an effective integration time of 1.6\,s. We chose the SUB512 (512\,$\times$\,32~pixels) subarray, which is wide enough in the cross-dispersion direction to allow for background subtraction using non-illuminated regions away from the trace. We used the NRSRAPID readout pattern and the S1600A1 (1.6\,×\,1.6) slit. We deliberately chose to saturate the detector between 1.1--1.8$\mu$m (using 6 groups instead of the maximum of 3 to avoid reaching saturation) to optimize the integration efficiency and the precision reached over the longer wavelengths (70\% more photons collected for each transit when factoring in the better duty cycle). This strategy enables us to increase our sensitivity to molecules that have significant opacity at $>2$$\mu$m. Both transit observations covered approximately 1.8 hours of pre-transit baseline and about 35 minutes after the transit.

\begin{figure*}
    \centering
    \includegraphics[width=0.95\textwidth]{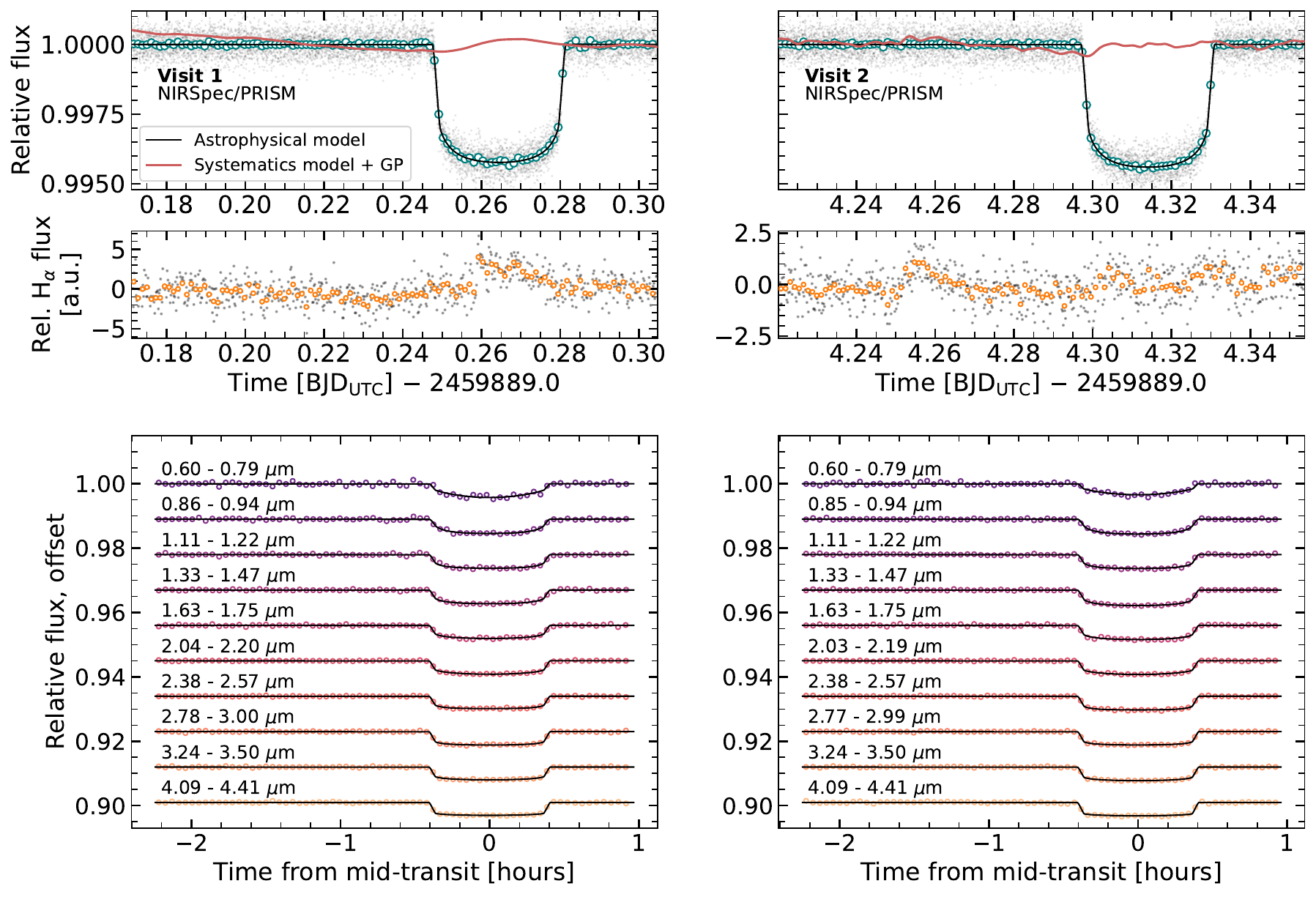}

    \caption{White and spectroscopic light-curve fitting results for the first (left) and second (right) NIRSpec/PRISM transit of TRAPPIST-1\,d. \textit{Top panels:} White light curves corrected by the best-fitting systematics models (gray points, and teal points are 60-points bins). The best-fit astronomical model (black) and systematics model including the GP component (red) are displayed on top of the data. \textit{Middle panels:} Median-subtracted integrated H$\alpha$ fluxes (in arbitrary units, multiplied by $10^{4}$) for visits 1 and 2, binned by a factor of 10 (gray points) and 60 (orange points). The timing and temporal shape of the astrophysical systematics captured by the GP strongly correlate with the H$_\alpha$ time series, suggesting that the flux variations are associated with small stellar flares during our observations. \textit{Bottom panels:} Ten sample systematics-corrected spectroscopic light curves, binned by a factor of 100 (colored points). The corresponding wavelength bins are labeled, and the best-fit astrophysical model is overlaid (black line). The light curves are offset relative to each other for clarity. The first spectroscopic bin (0.60-0.79 $\mu$m) is most affected by stellar activity and contains the H$\alpha$ line.}
    \label{fig:wlc_Halpha}
\end{figure*}

\subsection{Data reduction}\label{sec:eureka}

We perform the data reduction for both time series observations using the \texttt{jwst} package, the \texttt{Eureka!} data reduction pipeline \citep{bell_eureka_2022}, and custom routines. 

We first perform ramp fitting from the uncalibrated data products (standard Stage 1 pipeline) using the \texttt{jwst} pipeline, and custom routines. We follow the \texttt{jwst} pipeline's default settings, except for the few differences we outline here. We skip the jump detection step, which erroneously flags as outliers large fractions of the detector pixels, and instead rely on spatial and temporal outlier flagging at later stages in the data reduction. We perform on each image an equivalent of the reference pixel correction using the top and bottom 6 rows as reference pixels (with the spatial coordinate varying across rows), since no reference pixels are available on the NIRSpec detector. For each image, we calculate the median of odd and even rows among the defined reference pixels and subtract this median from all odd/even rows across the image. This reference pixel step is performed once before and once after the nonlinearity correction step. 

We use a custom nonlinearity correction step, where we perform the same steps as in Benneke et al. (submitted), where it was applied to other NIRSpec/PRISM time series observations of TRAPPIST-1. We start by using the default linearity correction coefficients on the median image, which leaves a residual curvature in the recorded flux as a function of the group number especially for highly illuminated pixels. We fit a quadratic function to this residual curvature as a function of the recorded flux in each row, which we use to correct all the images prior to ramp fitting (see Benneke et al. (submitted) for a detailed discussion). We add a custom routine for group-level background subtraction, where we record and subtract from each group image the median of the background pixels in each detector column for that group over all integrations, after discarding $3\sigma$ temporal outliers. 

The detector was deliberately saturated over a fraction of the set 6 groups per integration over a limited number of pixels from approximately 1.08 $\mu$m and 1.77 $\mu$m to optimize S/N and duty cycle in the rest of the spectrum. We use the default saturation detection routine from the \texttt{jwst} pipeline, and flag as saturated not only the affected pixels but the 3$\times$3 pixels centered on this pixel to account for charge migration. We tested a more aggressive treatment excluding from the ramp fitting not only the saturated group and subsequent groups, but also the previous group. We found that this did not impact our results and proceeded with the standard setup. Finally, we perform the ramp fitting step using the \texttt{jwst} pipeline algorithm. 

We use the Stage 2 pipeline from \texttt{Eureka!} (a wrapper around the \texttt{jwst} pipeline's Stage 2) to perform the final calibration steps prior to extraction (Stage 3). We follow the default steps, except that we skip the photometric calibration and the flat-fielding steps which are not required for transmission spectroscopy. We skip the \texttt{extract1d} step since we perform an independent spectral extraction using \texttt{Eureka!}.

We use \texttt{Eureka!} to perform a second iteration of background subtraction as well as optimal spectral extraction on each of the integrations, to convert the slope images into a time series of observed stellar spectra. We use rows 4 to 28 in the spatial direction and columns 14 to 495 in the dispersion direction. Pixels with any flags other than saturation flags are masked. The vertical position of the trace is identified by summing the image over detector columns and fitting a Gaussian profile to the resulting flux distribution. We perform background subtraction using the same routine as for Stage 1 to remove any remaining 1/$f$ noise. We rely on an additional step of background subtraction where we consider rows 7 pixels or more away from the trace, and where we discard pixels that are 10$\sigma$ outliers in time when constructing the background , in two outlier flagging iterations. Then, we loop through the odd and even rows of each group and subtract the mean value obtained from each set of rows. We then perform an optimal extraction weighted by the flux in the median image, with an aperture spanning seven rows above and below the trace position and record for each extracted spectrum in our time series the exact vertical position of the trace in the corresponding image and the trace width (standard deviation of the fitted Gaussian) in order to flag any instrumental systematics during the light-curve fitting step.

We then use \texttt{ExoTEP} \citep{benneke_spitzer_2017} to obtain the white and spectroscopic median-normalized light curves from the time series of stellar spectra. We also record potential shifts of the stellar spectrum in the dispersion direction using cross-correlation with the first extracted stellar spectrum, as well as the width of each cross-correlation peak. We perform further outlier removal prior to passing on the light curves to the fitting step: we clip data points that are 3$\sigma$ outliers in the spatial position of the trace on the detector, in the width of the Gaussian fitted to the trace, in the recorded spectrum shifts in the dispersion direction, and in the flux (identified using a running median with a window size of 20). 

We also record the integrated H$\alpha$ flux in each of our spectra, since jumps in this metric relative to its baseline have been found to track stellar flares on TRAPPIST-1 \citep{lim_atmospheric_2023,howard_characterizing_2023}. Since NIRSpec/PRISM does not have the high resolution required to resolve the H$\alpha$ line, we need to use a wider wavelength window compared to previous NIRISS/SOSS observations of TRAPPIST-1 \citep{lim_atmospheric_2023,RadicaT1c}, and measure the combined flux emitted by H$\alpha$ and a neighboring feature\footnote{Here we refer to the emission feature just shortwards of H$\alpha$ as seen in Fig. 8b of \citet{lim_atmospheric_2023}. Since this feature remained consistent between quiescent and flaring epochs \citep{lim_atmospheric_2023}, we conclude that the relative variations we are measuring indeed reflect the H$\alpha$ luminosity.} \citep{lim_atmospheric_2023}. We estimate the H$\alpha$ emission by fitting a straight line to the pseudo-continuum on either side of the blended spectral feature (0.005$\mu$m on each side of the 0.6460--0.6645 $\mu$m feature). This spectrum region is then normalized, and the continuum is subtracted out before integrating the flux contained over the considered wavelength range. This metric then corresponds to the H$\alpha$ flux, scaled by a factor of 10$^4$ in Figure \ref{fig:wlc_Halpha}.

\section{Transit light-curve fitting with ExoTEP}\label{sec:lc_analysis}

We fit the broadband and wavelength-dependent light curves using the \texttt{ExoTEP} pipeline (\citealp{benneke_spitzer_2017,benneke_sub-neptune_2019}; see Sections \ref{sec:broadband_lc} and \ref{sec:spec_lc}). We evaluate the impact of our choice of prescription for limb-darkening on the spectrum, and choose to adopt limb-darkening coefficients measured from an observation of another TRAPPIST-1 planet transit that happened during a time of relative stellar quiescence (no flares and no stellar contamination detected; see Section \ref{sec:LD}). We examine the H$\alpha$ luminosity time series and identify small increases refecting stellar flares (Figure \ref{fig:wlc_Halpha}). Such ``microflares'' were found to be frequent in a five-year \textit{Hubble space telescope} (HST) monitoring campaign of TRAPPIST-1 \citep{berardo_hubble_2025}. The flares occurred during the first transit, and in the pre-transit baseline as well as during the transit in the second visit. After testing several detrending methods to account for the wavelength-dependent effect of these small flares (e.g. Figure \ref{fig:raw_lightcurves}), we select a systematics model that captures a linear trend with time and models the remaining correlated noise as a Gaussian Process with a Matérn 3/2 kernel (see Section \ref{sec:GP}). 

We find that the planetary and orbital parameters inferred from the fit to visit 1 and visit 2 are in very good agreement (Figure \ref{fig:b_aRs}, Table \ref{table:wlc_param}) and use them to calculate derived parameters for TRAPPIST-1\,d (Table \ref{table:pla_star_param}). The planetary radius, semi-major axis, and inclination inferred from our two NIRSpec/PRISM transits (Table \ref{table:pla_star_param}) are all in excellent agreement with literature values obtained from fitting all the \textit{Spitzer}, ground-based, \textit{HST}, and \textit{K2} transit observations of the TRAPPIST-1 planets, with comparable uncertainties \citep{agol_refining_2021}. Our measured transit times (Table \ref{table:wlc_param}) also closely match (with a precision improved by a factor of 100) the predicted times of $\mathrm{BJD}=2459889.265414 \pm 0.003625$ and $2459893.314955 \pm 0.003632$ for the two visits, from the dynamical forecast informed by previous transit observations of the system \citep{agol_refining_2021}.

Interestingly, despite the good agreement in the inferred planetary parameters between both visits (Table \ref{table:wlc_param}, Figure \ref{fig:b_aRs}), we notice significant differences in shapes of the two transmission spectra (Figure \ref{fig:fitted_stctm_spectra}, and Table \ref{tab:merged_transm_spec}), which we attribute to the TLS signal varying over the span of 4 days between the first and the second observation, or approximately 1.2$\times$ the stellar rotation period of 3.3d (Table \ref{table:pla_star_param};\citealp{luger_seven-planet_2017}).

\section{Stellar surface modeling}\label{sec:stellar_ctm}

We leveraged the information from both the transit spectra and the out-of-transit stellar spectra to infer the distribution of dark and bright heterogeneities on the stellar surface during each visit. We assessed the visit-to-visit variability in the transit light source signature, and evaluated the extent to which the independent constraints from out- and in-transit analyses painted a consistent picture of TRAPPIST-1's surface at the epoch of the observations. We find that a quiet photosphere with a minor coverage ($\sim 10$\%) of unocculted bright heterogeneities produces the best match to our observations, while the properties of potential dark heterogeneities are unconstrained and their addition does not lead to a better fit quality.

\subsection{Methodology}\label{sec:stctm_methods_and_TLSsub_spectrum}
We obtain flux-calibrated out-of-transit stellar spectra using a modified version of the data reduction described previously. Since this is the first description of flux-calibrated NIRSpec/PRISM data to our knowledge, we describe how we customized the \texttt{Eureka!} pipeline for this purpose (see Section \ref{sec:fluxcal}). In order to interpret each visit's out-of-transit flux-calibrated spectra in terms of the stellar surface properties, we present a new out-of-transit spectrum retrieval sub-module, \texttt{exotune} (EXoplanet hOst ouT-of-transit spectrUm fittiNg framEwork), implemented into the open-source stellar contamination modeling code \texttt{stctm}\footnote{\url{https://github.com/cpiaulet/stctm}}  (\citealp{Piaulet_Ghorayeb_2024,stctm_zenodo_temp}, Piaulet-Ghorayeb, subm.). Our procedure provides a statistically robust way to obtain posterior distributions on relevant stellar photosphere and heterogeneity properties within a Bayesian framework rather than performing a grid search for the highest-likelihood model (e.g. \citealp{wakeford_disentangling_2019,moran_high_2023,davoudi_updated_2024}). In contrast to previous work (e.g. \citealp{RadicaT1c}), our out-of-transit stellar spectrum retrievals are fully parallelized, enabling fast computation even for the large number of data points expected from spectra obtained with JWST NIRISS SOSS or NIRSpec. The methodology applied to infer the properties of the stellar surface is described in Section \ref{sec:oot_spectrum_methods}. This cross-check serves as further validation of the stellar surface heterogeneity properties we infer from the standard \texttt{stctm} fit to the impact of TLS on the transmission spectra of TRAPPIST-1\,d (see Section \ref{sec:TLS_methods}).

\begin{figure*}
    \centering
    \includegraphics[width=0.99\linewidth]{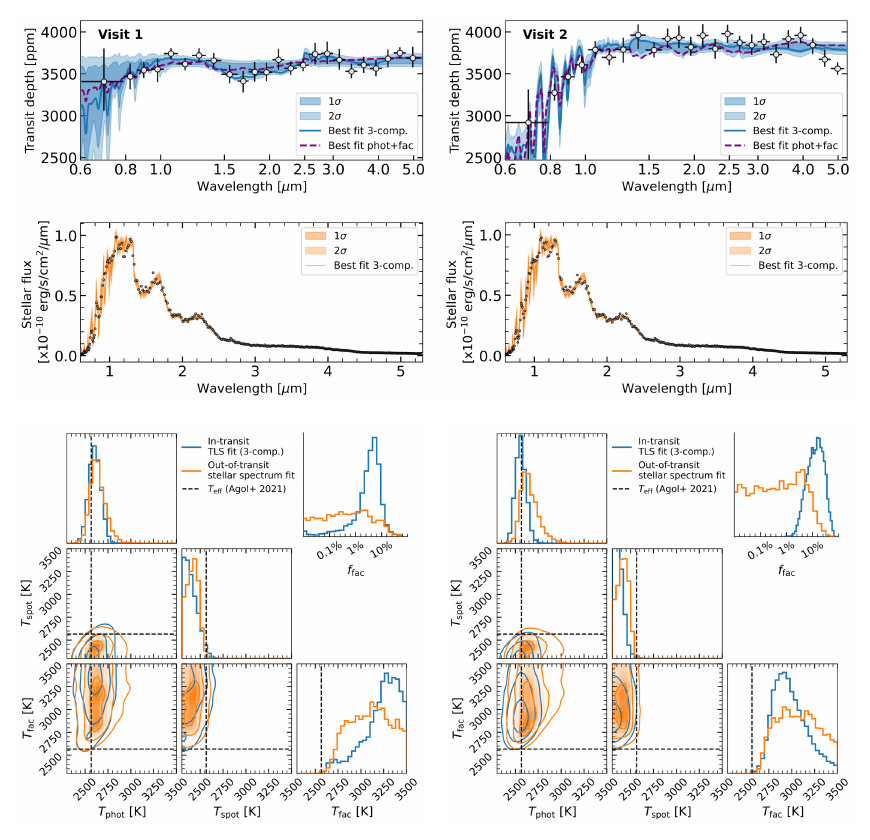}
    \vspace{5mm}
    \vspace{-5mm}\caption{
    Results from the retrievals performed on the  transmission spectra of TRAPPIST-1\,d and out-of-transit stellar spectra of TRAPPIST-1 with \texttt{stctm}. The retrievals assume that the spectra can be fully explained by a three-component model including the stellar photosphere, and both cooler (`spot') and brighter (`facula') heterogeneity components, with no wavelength-dependent atmosphere signature in the transmission spectrum. \textit{Top panels:} Measured transmission spectra  (black points) and  best-fit models (blue) along with the 1- and 2$\sigma$ retrieved contours on the stellar spectrum (color shadings). The dashed purple line corresponds to the best-fit model in a two-component retrieval without spots. \textit{Second panels:} Same thing, for the retrieval performed on the out-of-transit stellar spectrum, in shades of orange. \textit{Bottom panels:} Joint and marginalized posterior probability distributions on the temperatures of the photosphere, spot, and facula component from both types of retrievals, with 1, 2, and 3$\sigma$ contours. The independently-determined stellar effective temperature from \citet{agol_refining_2021} is shown with dashed lines. We display in the upper right corner the marginalized posterior distribution on the covering fraction of the facula component (the spot covering fraction was unconstrained in all our retrievals, see Table \ref{tab:retrieved_stctm_params}). We find good agreement between the stellar properties inferred from the TLS and out-of-transit spectrum retrievals.}
    \label{fig:fitted_stctm_spectra}
\end{figure*}


We use the results from the \texttt{stctm} retrieval to create ``TLS-corrected'' versions of the transmission spectra ($D_\mathrm{\lambda,corr}$) for each visit. These spectra are obtained by dividing each measured transmission spectrum $D_\mathrm{\lambda,meas}$ by the best-fitting model from the three-component TLS-only retrieval ($D_\mathrm{\lambda,TLS}$), and multiplying the ratio of these spectra by the value of the bare-rock transit depth ($D_\mathrm{scale, TLS}$; see Section \ref{sec:TLS_methods})) corresponding to the best-fit model (3587 ppm for visit 1, 3658 ppm for visit 2):

\begin{equation}
    D_\mathrm{\lambda,corr}=\frac{D_\mathrm{\lambda,meas}}{D_\mathrm{\lambda,TLS}} \times D_\mathrm{scale, TLS}.
\end{equation}

The correction aims at effectively dividing the measured transmission by the stellar contamination factor $\epsilon_\mathrm{\lambda,het}$, since $D_\mathrm{\lambda,TLS}$ is understood in the \texttt{stctm} modeling formalism as:
\begin{equation}
    D_\mathrm{\lambda,TLS} = D_\mathrm{scale, TLS} \times \epsilon_\mathrm{\lambda,het}.
\end{equation}
\noindent where $\epsilon_\mathrm{\lambda,het}$ is a factor representing the contamination by stellar heterogeneities (see \citealp{Piaulet_Ghorayeb_2024}, eq. C1).

We leverage these spectra to perform atmospheric retrievals using the ``sequential'' approach, where the stellar contamination contribution is first divided out of each visit's transmission spectrum prior to combining them for the atmospheric retrievals, and assess the impact of joint atmosphere and TLS marginalization on our sensitivity to atmosphere signatures. When displaying the combined visit 1+2 ``TLS-corrected'' transmission spectrum, we shift the visit 2 spectrum by the difference between the bare-rock transit depths of visits 1 and 2 and then show the error-weighted average of these two spectra. That spectrum is only used for visualization purposes, as we use directly the individual visit 1 and 2 TLS-corrected spectra in our SCARLET retrievals (while marginalizing over a visit-to-visit offset; see Section \ref{sec:atm_modeling}).

\subsection{Inferred stellar surface properties}

We find that both spectra can be fully explained by stellar contamination (Figure \ref{fig:fitted_stctm_spectra}), and find tentative evidence for the presence of bright surface heterogeneity regions (faculae) at the level of 1.8$\sigma$ for visit 1, and 2.8$\sigma$ for visit 2 while the addition of colder spots to the model is not favored (Section \ref{sec:TLS_results}). The TLS and out-of-transit spectrum retrievals both yield very consistent posterior constraints on the retrieved stellar surface properties, despite the uniform prior adopted for $T_\mathrm{phot}$ in the \texttt{exotune} retrievals (Figure \ref{fig:fitted_stctm_spectra}). Further, the out-of-transit \texttt{exotune} retrievals provide an independent constraint on the photospheric temperature (Table \ref{tab:retrieved_stctm_params}), and place a stricter upper limit of about 10\% on the covering fraction of the hotter `faculae' spectral component (Table \ref{tab:retrieved_stctm_params}). 

We note that model fidelity is a pervasive issue affecting the interpretation of M dwarf spectra, and our study is not immune to these biases (see e.g. \citealp{Rackham_SAG21_2023,deWit_roadmap_2023, davoudi_updated_2024,jahandar_chemical_2024}). We find that the wider wavelength coverage of NIRSpec/PRISM is helpful in this regard as it allows us to capture the Rayleigh-Jeans tail of the stellar SED and to be less sensitive to modeling issues at the shortest wavelengths that plagued NIRISS/SOSS interpretation \citep{lim_atmospheric_2023,RadicaT1c}. We also compared our spectrum to a SPHINX, rather than a PHOENIX, model, and found that the mismatch is at least comparable to what PHOENIX models display for the nominal literature stellar parameters (Figure \ref{fig:oot_phoenix_sphinx}). We discuss these caveats in more detail in the Appendix (Section \ref{sec:caveats_oot_TLS}). 

Finally, our approach does not account for the wavelength-dependent signatures of flare signals (see Figure \ref{fig:raw_lightcurves}), which may affect the out-of-transit and transmission spectra. Instead, here, we adopt the approach of avoiding flare epochs when constructing the out-of-transit spectrum, and retrieving the effect of the TLS effect on transmission spectra that are effectively affected by both TLS and flares (although some of the impact of flares may be taken out by the GP fitting). A more extensive investigation of this impact (Lim et al., in prep.) found that flares modeled under the simplifying assumption of a time-decaying blackbody hotter than the photosphere may mimick the spectral signatures of unocculted faculae, but this conclusion only holds if no GP is fitted to the associated correlated noise, while we model the wavelength-dependent flare signature with a Matérn 3/2 GP in this work. Applying a GP to the light curves should remove this impact in theory, and might even introduce a spurious spot-like \textit{increase} in transit depths at the shortest wavelengths where the flare signal is poorly modeled by the GP, which is not detected in our case (Lim et al., in prep.). A blackbody function is not a perfect match to the near-infrared flare spectra of TRAPPIST-1, \citep{howard_characterizing_2023}. Instead, flare spectra are expected to exhibit short- and long-wavelength components that evolve on different timescales (see e.g. \citealp{kowalski_timeresolved_2013}). We defer a full exploration of these effects to future work and encourage the community to invest effort into understanding the impact of both the TLS effect, flares, and their wavelength-dependent signatures depending on the amplitude and timing of the flare relative to the transit in order to better assess their impact on small-planet atmosphere exploration.



\section{Complementarity of in-transit and out-of-transit retrievals} \label{sec:complementarity_oot_TLS}

The main conclusion from all the retrievals we performed is consistent across in- and out-of-transit analyses: a photosphere component and a minor facula component provide a good match to both the out-of-transit stellar spectrum, and the imprint of the TLS effect on the transmission spectrum (Figure \ref{fig:fitted_stctm_spectra}). We find that the main strengths of out-of-transit spectral retrievals lies in their sensitivity to the temperature of the photospheric component and their independence from planetary atmospheres, while the TLS retrievals can more accurately detect spectral components with low covering fractions because of the greater fractional impact they have on the transmission spectrum.

\paragraph{Strengths of out-of-transit spectrum retrievals}
The transmission spectrum can only probe the \textit{ratios} of the photosphere to heterogeneity components through the TLS effect, and therefore relies entirely on a literature prior for the temperature of the quiet photosphere $T_\mathrm{phot}$. Meanwhile, the out-of-transit spectrum retrievals is directly sensitive to $T_\mathrm{phot}$. For both visits, our \texttt{exotune} retrievals find photosphere temperatures in perfect agreement with independent determinations in the literature (Figure \ref{fig:fitted_stctm_spectra}, Tables \ref{table:pla_star_param} and \ref{tab:retrieved_stctm_params}).
We also obtain a stricter upper limit on the contribution of the faculae component to the stellar spectrum ($f_\mathrm{fac}$) compared to TLS retrievals alone thanks to the capacity to leverage the entire spectral SED. This upper limit likely reflects an overall constraint on the range of stellar temperatures that can explain TRAPPIST-1's NIRSpec/PRISM spectrum; a larger contribution from a hot component would lead to noticeable changes in both spectral shape and intensity.

\paragraph{Strengths of transit light source effect retrievals}
While the \texttt{exotune} retrievals can directly probe the emission from the stellar surface, they lack the spatial information of how the stellar surface outside of the transit chord looks relative to that underlying the path of the planet's transit. In our case, we do not detect any spot or facula crossing event, suggesting that the assumption of the planet transiting over the quiet photosphere is warranted here. Despite the limited sensitivity to \textit{temperature} contrasts, a distinct advantage of the TLS retrievals relative to their out-of-transit counterparts is their ability to tease out very minor stellar spectral components. This translates to a lower limit on $f_\mathrm{fac}$ from the TLS retrieval in the case of the second visit, driven by the $\sim 1000$\,ppm transit depth variations observed at short wavelengths. Therefore, TLS retrievals can guide the setup of out-of-transit retrievals thanks to their sensitivity to smaller heterogeneity contributions which can be degenerate (between e.g. a large fraction of spots vs. a cooler photosphere) in the exotune fit. In our case, the TLS retrievals motivated the exploration of at least two spectral components (photosphere+faculae) to interpret the out-of-transit spectrum.
 

Importantly, while both methods can yield posterior distributions on the same parameters, the fitted quantities are fundamentally different: the TLS fit probes the ratios of the spectra for the different stellar surface components (but can be biased by potential planetary atmosphere contributions), while the out-of-transit fit is sensitive to the absolute SED. This makes these approaches complementary, and the important of accurate stellar modeling for inferences about planetary atmospheres motivates us to argue for performing this cross-check for all transiting planet observations where a stellar spectrum is available. 
Finally, the results from out-of-transit stellar spectrum retrievals can be used to place informed priors on the TLS components of joint atmosphere-TLS retrievals. Here, for instance, both the TLS and stellar spectrum retrievals support the conclusion that there is no evidence for a difference in the representative $\log g$ of the photosphere and heterogeneity components (Table \ref{tab:retrieved_stctm_params}), a conclusion we leverage in the setup of our atmosphere retrievals.

\section{Atmosphere modeling} \label{sec:atm_modeling}

\subsection{SCARLET atmosphere retrievals}
We use SCARLET \citep{benneke_distinguishing_2012,benneke_characterizing_2013,benneke_strict_2015,benneke_sub-neptune_2019, benneke_water_2019, pelletier_where_2021, piaulet_evidence_2023} to perform both atmosphere-only (Section \ref{sec:sequential_SCARLET}) and joint atmosphere-stellar contamination (Section \ref{sec:joint_star_planet_retrieval}) retrievals on the NIRSpec/PRISM spectra of TRAPPIST-1\,d. We explore for the first time the impact of the ``sequential'' retrievals performed in previous works on the TRAPPIST-1 planets (\citealp{lim_atmospheric_2023} for TRAPPIST-1 b, Benneke et al., in review for TRAPPIST-1 g) on the atmospheric inferences. In the ``sequential'' retrieval approach, retrievals are performed on the TLS-corrected (visit 1 + visit 2) transmission spectrum (see Section \ref{sec:stctm_methods_and_TLSsub_spectrum}). We compare the results from this two-step process to the inferences from a joint retrieval where atmospheric properties are shared across both visits, while stellar heterogeneity parameters are fitted independently for each visit. We run three types of retrievals: (1) assuming atmospheres with H$_2$/He as the background molecules, and freely fitted abundances for other species (prior favoring low-mean molecular weight atmospheres); (2) assuming pure-composition atmospheres made of either CH$_4$, H$_2$O, CO$_2$, CO, or NH$_3$, all of which our NIRSpec/PRISM wavelength range is sensitive to; and (3) exploring atmospheres where N$_2$ is mixed with an IR-absorbing species such as CH$_4$, CO$_2$ or H$_2$O, which allows for a comparison with solar system terrestrial compositions (see details of the implementation in Section \ref{sec:atm_model_methods}). For the joint retrievals, we assume that the $\log g$ of the heterogeneities and photosphere are identical (based on the results from the stellar surface modeling step) but allow for both visit-specific spot and facula components (see Section \ref{sec:joint_star_planet_retrieval}) since spot signatures could mimick atmospheric absorption signals and are not independently ruled out by the out-of-transit spectrum analysis. Finally, we assess the extent to which the limits we place on potential atmospheric thicknesses are impacted by the refraction of stellar light away from our line of sight (see Section \ref{sec:refraction}) for two edge-case scenarios among the cases we explore (since the atmospheric mean molecular weight correlates with the refractive index): a water-dominated (10 bar H$_2$O, 1 bar N$_2$) composition, and a pure-CO$_2$ (10 bar CO$_2$) composition.

\subsection{Constraints on potential planetary atmospheres}\label{sec:constraints_atm}
\begin{figure}
    \centering
    \includegraphics[width=0.95\linewidth]{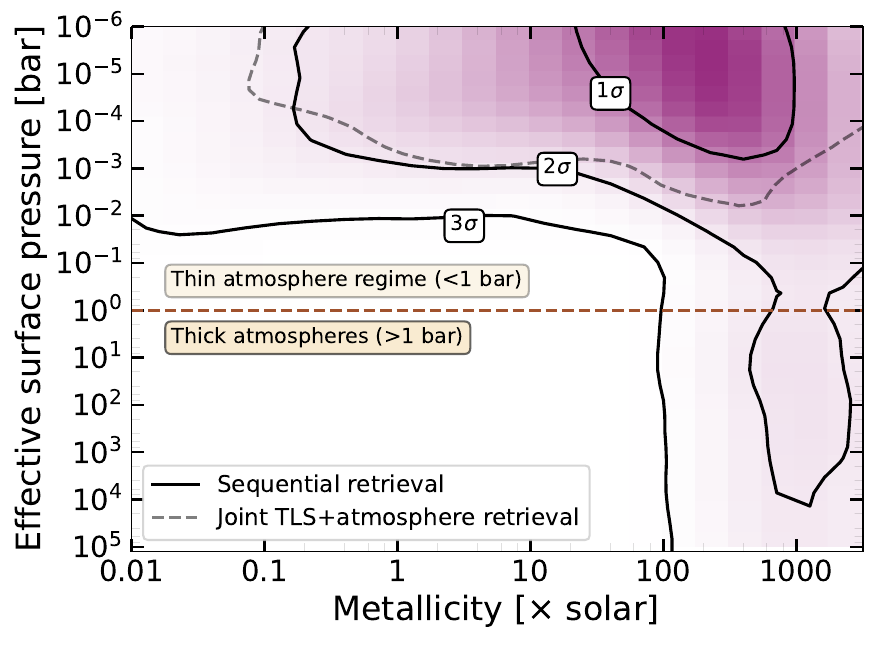}
    \vspace{5mm}
    \vspace{-5mm}\caption{Constraints on the effective surface pressure of a potential atmosphere (with H$_2$/He assumed to be the background ``filler gas'') as a function of atmospheric metallicity. We show the joint posterior probability distribution on the effective surface pressure (of either a bare-rock surface or a gray cloud deck) and the atmospheric metallicity calculated from the posterior distributions on the H$_2$/He, H$_2$O, CO, CO$_2$, CH$_4$, N$_2$, NH$_3$ and SO$_2$ abundances. The purple shading indicates high-probability regions, and the contour lines trace the 68\%, 95\% and 99.7\% confidence regions for the sequential retrieval (solid) and the 95\% contour in the joint retrieval case (dashed). At the 3$\sigma$ level, our spectrum excludes hydrogen-dominated atmospheres down to 0.01 bar effective pressures, and up to 100$\times$ solar metallicity, while some high mean molecular weight atmospheres remain compatible with the spectrum. }
    \label{fig:met_psurf}
\end{figure}

\begin{figure*}
    \centering
    \includegraphics[width=0.9\linewidth]{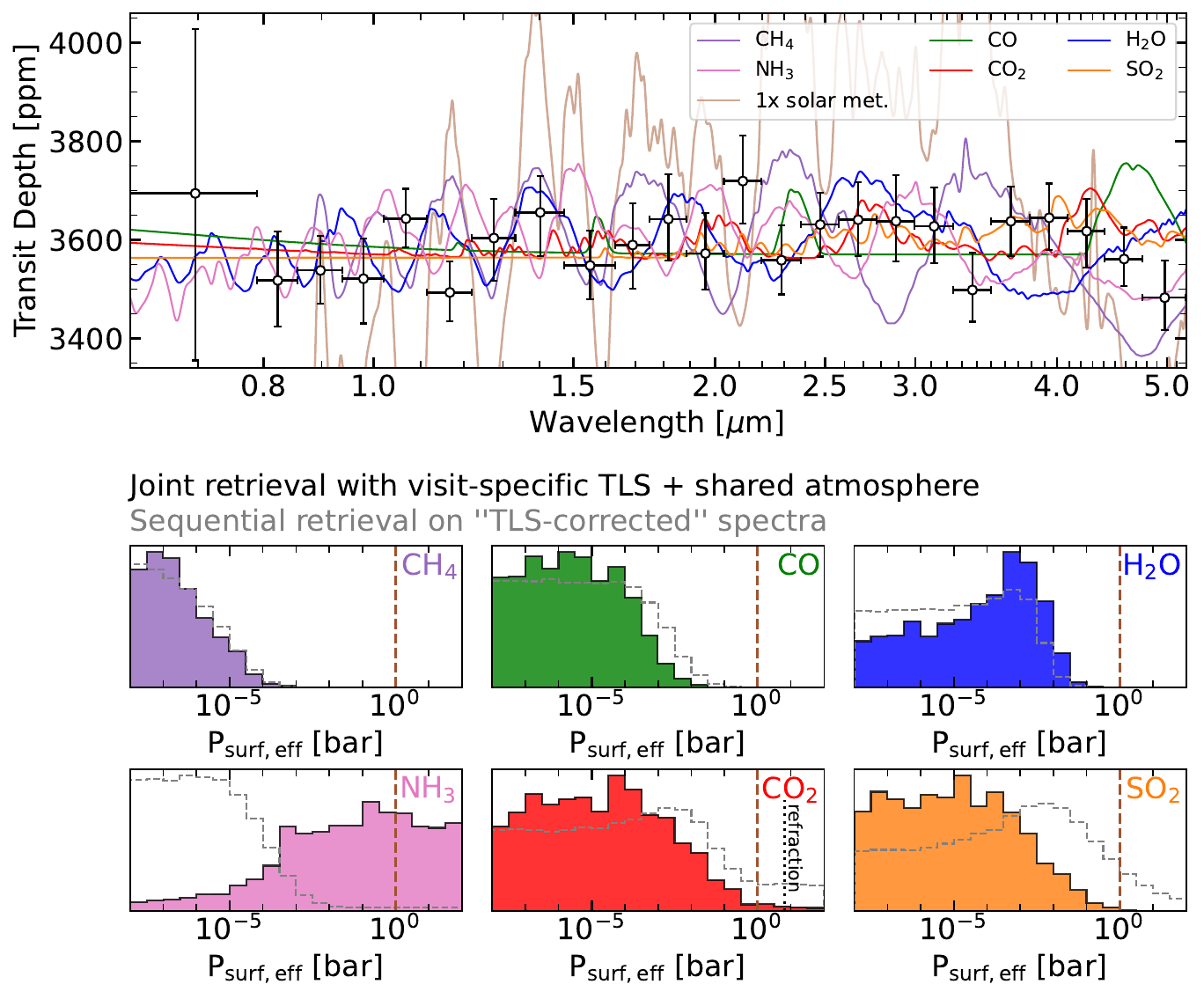}
    \vspace{5mm}
    \vspace{-5mm}\caption{Constraints on the effective surface pressure of a potential high mean molecular weight atmosphere on TRAPPIST-1\,d. \textit{Top panel:} The combined Visit 1+2 transmission spectrum is shown after stellar contamination correction (black points, spectrum used for the sequential retrievals). We show representative forward models for a cloud-free solar-composition atmosphere and 1-bar atmospheres} made of 100\% CH$_4$, CO, H$_2$O, NH$_3$, CO$_2$, or SO$_2$, smoothed to a resolving power of R=100. \textit{Bottom panels:} Marginalized posterior distributions on the effective surface pressure of CH$_4$, CO, H$_2$O, NH$_3$, CO$_2$, and SO$_2$ (gray: sequential fit; solid filled distributions: joint fit with stellar contamination). The dashed vertical lines highlight the location of a 1-bar atmosphere. For the 100\% CO$_2$ case, we show as a dotted line the refraction limit for a 10-bar atmosphere.
    \label{fig:atmosphere_molbymol}
\end{figure*}


We first perform a standard well-mixed multi-gas atmosphere retrieval with H$_2$/He as the background gases, and either the ``sequential'' or ``joint retrieval'' approach to treat the TLS effect. This test allows us to assess the compatibility of our observations with different atmospheric mean molecular weights (from hydrogen-rich to metal-enriched atmospheres) across a wide range of atmospheric thicknesses (above a thick cloud deck or a rocky surface). The shape of the joint distribution on the atmospheric metallicity (Figure \ref{fig:met_psurf}; metallicity calculated from the abundances of all the gases included: H$_2$, He, H$_2$O, CH$_4$, CO, CO$_2$, N$_2$, SO$_2$ and NH$_3$) reflects the fact that our transmission spectrum does not show detectable planetary absorption features. The precision we reach enables us to discard with a high level of confidence cloud-free hydrogen-dominated atmospheres (see also Figure \ref{fig:atmosphere_molbymol}), in agreement with previous studies \citep{de_wit_atmospheric_2018,ducrot_0.8-4.5mum_2018,ducrot_trappist-1_2020}. However, for conditions rich in high mean molecular weight volatiles, thick cloud-free atmospheres remain compatible with the spectrum. This conclusion is likely driven by our inclusion of NH$_3$, SO$_2$ and CO$_2$ in the model (Figure \ref{fig:atmosphere_molbymol}, see Section \ref{sec:constraints_atm}). 

This motivates us to further explore the potential chemical makeup of high mean molecular weight atmospheres compatible with the observed spectrum, and our sensitivity to the presence of specific infrared absorbers. There, we turn to the results of our single-molecule retrievals, where we fit for effective surface pressures of an atmosphere made of 100\% H$_2$O, CH$_4$, CO, SO$_2$, CO$_2$ or NH$_3$. None of the molecules that we probed were detected (Figure \ref{fig:atmosphere_molbymol}) and we find that even thin, high mean molecular weight atmospheres of 1 bar are excluded with better than 95\% confidence for 100\% CH$_4$, CO, H$_2$O, and CO$_2$ atmospheres in the most conservative case of a joint retrieval for the TLS and atmosphere contributions. For NH$_3$, however, we only obtain a meaningful upper limit in the sequential fit (Table \ref{tab:atm_table}, Figure \ref{fig:atmosphere_molbymol}) which vanishes when we marginalize over all possible TLS realizations. Finally, SO$_2$ and CO$_2$ only have a few very small amplitude features within reach of NIRSpec/PRISM, especially in these conservative 100\% composition cases which lead to the overall highest atmospheric mean molecular weights of all the tests we ran. For these two molecules, our results depend on the assumption on the TLS treatment, and are more sensitive to the details of how the stellar contamination correction is performed. This difference is likely driven by the single-molecule modeling choice, since these two species only have a few localized absorption bands over the wavelength range of NIRSpec/PRISM.
 
We also run retrievals under the assumption that a large part of the atmospheric mass may be contained in a high mean molecular weight species without significant infrared absorption features, such as N$_2$ (as is the case on Earth) which only has a weak feature from collision-induced absorption at $\sim 2.2$$\mu$m. For these retrievals (focusing on the results obtained from the joint TLS+atmosphere fits), we obtain joint constraints on the partial pressures of an atmosphere made of N$_2$ and one of the IR absorbers CH$_4$, H$_2$O, or, CO$_2$, allowing us to compare our constraints with the compositions of solar system terrestrial atmospheres (Figure \ref{fig:partialps}). We obtain upper limits for all three molecules, reaching the sensitivity required to exclude abundances greater than 500 ppb for CH$_4$, 6000 ppb for CO$_2$, and 1412 ppm for H$_2$O in a 1 bar N$_2$-background atmosphere (or 8.9 ppm, 158 ppm, and 0.1\% respectively for a thin 0.1 bar N$_2$-background atmosphere) within the 95\% confidence interval. 

Although atmospheric refraction is crucial to consider for TRAPPIST-1\,d since it orbits a small late M dwarf, we find that it does not impact the upper limits we obtain from the retrievals. Specifically, we find that in the thick 10-bar pure-CO$_2$ (water-rich) scenario, refraction cuts off the contribution from layers deeper than 6.61 (1.67) bar. For the CO$_2$ atmosphere case, this corresponds to pressures higher than the upper limits we place (Figure \ref{fig:atmosphere_molbymol}). For the N$_2$-H$_2$O case, this cutoff only affects the deep, 10 bar H$_2$O+1 bar N$_2$ case (where the signal from pressures deeper than $\sim 2$ bar would be cut off), but the thinner atmosphere scenarios where most of our posterior lies are largely unaffected by this effect (Figure \ref{fig:partialps}). Altogether, these results underscore the sensitivity of our TRAPPIST-1\,d spectrum to high mean molecular weight atmospheres with surface pressures even thinner than the Earth, enabling detailed comparisons to solar system terrestrial atmospheres (see Section \ref{sec:high_mmw}.

\subsection{Impact of stellar contamination on the inferences}\label{sec:impact_TLS_on_scarlet}

Our modeling setup enables us to assess the impact of the retrieval assumptions on the sensitivity of our high-precision transmission spectrum to a variety of high mean molecular weight compositions. This modeling exercise opens up the possibility to uniquely assess the impact of the unknown stellar contamination on the inferences that can be made about the atmosphere, which matters substantially for TRAPPIST-1\,d as both visits are seemingly affected by stellar contamination at significantly different degrees (Figure \ref{fig:fitted_stctm_spectra}).

We find that, for TRAPPIST-1\,d, the limits we obtain on CH$_4$ and CO are not substantially altered by the joint fitting of stellar and planetary contributions (Figure \ref{fig:atmosphere_molbymol}). However, constraints on the presence of thin ($<1$bar) atmospheres are weakened for the pure-H$_2$O case (with an upper limit moving to pressures a factor of 2 higher), and essentially vanish for the pure-NH$_3$ case when visit-dependent stellar contamination is marginalized over. This behavior is largely expected for water, as we demonstrated with the same retrieval framework in \citet{RadicaT1c} for the case of TRAPPIST-1 c. This is in line with the theory of the transit light source effect \citep[e.g.,][]{rackham_transit_2018}, as specifically unocculted bright regions can imprint inverted water features on the spectrum that may mask planetary absorption signatures. Stellar contamination features, however, do not typically create NH$_3$-like signatures. For ammonia, our results can be traced back to the relatively smaller amplitude of NH$_3$ features compared to H$_2$O and CH$_4$ from the molecular cross-sections (for similarly high molecular masses), and to the fact that the stellar contamination model has enough flexibility to alter the shape of each visit's transmission spectrum in a way that does not rule out NH$_3$ features in a joint retrieval setup. The cases of SO$_2$ and CO$_2$ are trickier to explain to the extent that the theoretically more conservative joint retrieval setup yields stricter constraints than the sequential retrieval. It can, however, be understood in the context that the CO$_2$ and SO$_2$ features are much more localized than their CH$_4$, H$_2$O or NH$_3$ counterparts: in the 100\% composition case, the model spectrum is essentially flat throughout most of the NIRSpec/PRISM wavelength range with a few absorption features that can be cut off by clouds. If these happen to line up with small deviations in the TLS-corrected spectra (Figure \ref{fig:atmosphere_molbymol}), the constraints may be wider in the sequential case than in the joint retrieval case, where the parameters of the TLS contribution are optimized to yield a flat spectrum at least up to about 1.5$\mu$m. This explanation is in line with the similarity of the results we obtain for CO (to a lesser extent), which also has only a few features between 0.6 and 5.2 $\mu$m. Still, none of the compositions allowed by the posterior probability distribution from the sequential fit case are formally significantly ruled out in the joint fit setup, which is consistent with the sequential fit corresponding to a smaller, but nested, prior probability space.
 




\section{Discussion of remaining atmosphere prospects} \label{sec:disc}



Our precise transmission spectrum can be fully explained by stellar contamination alone, and therefore enables us to rule out cloud-free or thick atmosphere scenarios across a wide range of potential atmospheric metallicities (Figure \ref{fig:met_psurf}). We explore the plausibility of the atmospheric scenarios that cannot be ruled out from the transmission spectrum alone: a thin or cloudy hydrogen-rich atmosphere, or a high mean molecular weight atmosphere. We discuss our sensitivity to atmospheres similar to solar system terrestrials, the impact of 3D atmosphere circulation and cloud condensation on the detectability of a water-rich atmosphere, as well as the implications of a bare-rock scenario for the formation of TRAPPIST-1\,d and for atmosphere prospects on the colder planets TRAPPIST-1 e, f, and g.


\subsection{Ruling out thick hydrogen-rich atmospheres} \label{sec:rule_out_h2}

The joint effective surface pressure-atmospheric metallicity posterior obtained for TRAPPIST-1\,d (Figure \ref{fig:met_psurf}) leaves only two possibilities for H$_2$-rich, low-metallicity atmospheres: either we are seeing all the way down to the planetary surface, and TRAPPIST-1\,d has an extremely thin (less than 0.01 bar at the 3$\sigma$ level) hydrogen-rich atmosphere, or the conditions are met for clouds or hazes to form at pressures of $\lesssim0.01$ bar.

The cloudy case is disfavored by both modeling and laboratory work. Even if atmosphere observations alone may not rule out cloudy hydrogen-rich atmospheres, high-altitude clouds or hazes are extremely unlikely to form at low pressures, muting spectral features for TRAPPIST-1\,d \citep{moran_limits_2018,morley_thermal_2015}. More generally for H$_2$-rich atmospheres, we have to consider both sources and sinks of the atmospheric H$_2$. From a formation standpoint, only small amounts of H$_2$ could have been accreted by TRAPPIST-1\,d from the protoplanetary disk during the formation of the rapidly migrating planet chain \citep{hori_trappist-1_2020}. Further, any primordial hydrogen was likely lost over time due to intense XUV irradiation from the host star during its lengthy pre-main-sequence phase, and continued atmospheric loss over the system's lifetime \citep{hori_trappist-1_2020,turbet_review_2020}. Although H$_2$ can also be provided by outgassing from a mantle with a reduced composition, modeling of the outgassing rate found it to be too low to replenish a hydrogen-dominated atmosphere relative to the escape rate \citep{hu_narrow_2023}. 

Therefore, we conclude that (1) thick cloud-free hydrogen-rich atmospheres are ruled out by our transmission spectrum, while (2) thin H$_2$-rich alternatives are strongly disfavored when considering TRAPPIST-1\,d in the context of its formation and evolution under the stellar irradiation, while (3) high-altitude clouds or hazes are not expected to form on TRAPPIST-1\,d if it has a low-metallicity atmosphere.

\begin{figure*}
    \centering
    \includegraphics[width=0.9\linewidth]{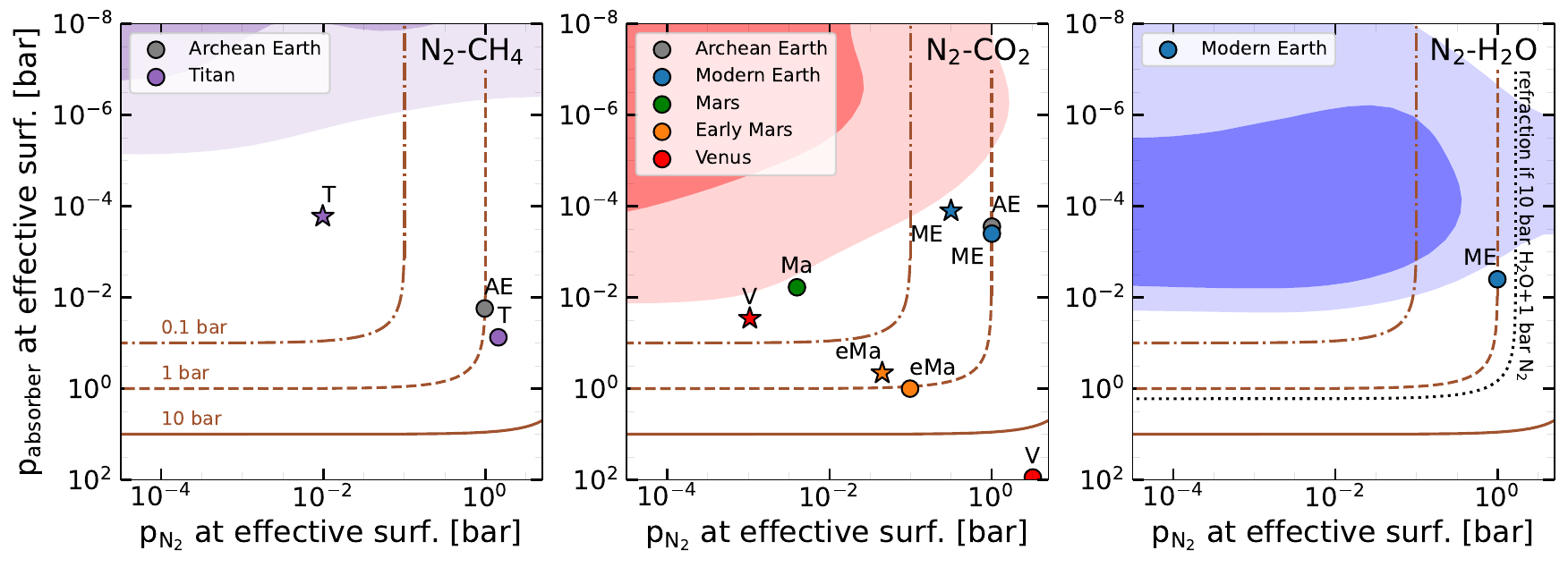}
    \vspace{5mm}
    \vspace{-5mm}\caption{Joint posterior constraints on the surface partial pressure of key atmospheric absorbers for atmospheric scenarios similar to solar system terrestrials, with N$_2$ as the background gas. The color shadings correspond to the posterior probability mass, and the contours highlight the 68\% and 95\% confidence regions from the joint retrieval in each case. The absorber in the N$_2$-background atmosphere is labeled in the top right corner of each panel. The dashed-dotted, dashed, and solid sienna lines correspond to an atmospheric surface pressure of 0.1, 1, and 10 bar. We illustrate where typical models of Archean Earth (AE), Modern Earth (ME), and compositions for Titan (T), early Mars (eMa), Mars (Ma), and Venus (V; \citealp{encrenaz_composition_2018}) approximately lie relative to our constraints for TRAPPIST-1\,d. Star markers correspond to specific cloudy cases, where the planet is modeled as either cloudy or cloud-free (for Titan, Modern Earth, and Venus). For the N$_2$-H$_2$O case, we show as a black dotted line the limit of pressures that would be cut off by refraction in a 11 bar atmosphere (10 bar H$_2$O+1 bar N$_2$). For thinner atmospheres, the impact of refraction on atmosphere detectability is even weaker.}
    \label{fig:partialps}
\end{figure*}

\subsection{Plausibility of a high mean molecular weight atmosphere} \label{sec:high_mmw}

We find that atmospheres thicker than 1 bar dominated by CH$_4$, CO, or H$_2$O are excluded with high confidence (Figure \ref{fig:atmosphere_molbymol}, Table \ref{tab:atm_table}, Figure \ref{fig:partialps}). The absence of CH$_4$ is not surprising, as methane is expected to be quickly depleted under the irradiation conditions of TRAPPIST-1\,d, from its photodissociation by incoming stellar photons and the formation of organic hazes \citep{turbet_modeling_2018,turbet_review_2020}, except in the presence of a very strong source, e.g., biotic methanogenesis \citep{krissansen-totton_disequilibrium_2018,meadows_feasibility_2023}. However, CO is a potential outcome for outgassed atmospheres from mantle-atmosphere geochemistry (e.g., \citealt{liggins_growth_2021}), although its abundance can vary widely depending on the redox state of the mantle, and the mantle melt fraction \citep{bower_retention_2022}. 

The constraints on pure-H$_2$O atmospheres are more nuanced. The upper limit on the effective surface pressure for H$_2$O is similarly constraining to those of CH$_4$ and CO atmospheres but may reflect the formation of high-altitude water clouds rather than a thin H$_2$O atmosphere (see next subsection). 

Although marginally disfavored by the transmission spectrum, atmospheres dominated by NH$_3$, CO$_2$, or SO$_2$ have lower-amplitude features that cannot be ruled out given our error bars (Figure \ref{fig:atmosphere_molbymol}). However, NH$_3$ is not expected to be a stable atmospheric component, as it is rapidly depleted by photodissociation, similarly to methane \citep{turbet_modeling_2018}. Further, interior degassing can produce SO$_2$, but in lower amounts than CO$_2$, making the SO$_2$-only atmosphere an unlikely edge case \citep{liggins_growth_2021,bower_retention_2022}. A mixed CO$_2$-SO$_2$ atmosphere, indicative of a highly oxidized mantle, remains within the realm of possibility -- although we note that the majority of the posterior mass for CO$_2$ still lies at very low effective surface pressures.

We then turn to comparing the constraints we obtain on high mean molecular weight atmospheres where an infrared absorber is mixed with N$_2$ to constrain further the plausible composition of an atmosphere on TRAPPIST-1\,d (Figure \ref{fig:partialps}). Comparing with the atmospheres of solar system terrestrial bodies, we find that our methane constraints enable us to confidently exclude Titan-like atmospheric compositions \citep{charnay_titan_2014} with 1.6\% CH$_4$ in a N$_2$ background atmosphere at better than 3$\sigma$. Even when considering the ``cutoff'' effect of Titan's haze layer on the transmission spectrum at 0.01 bar (``cloudy Titan'' case; Figure \ref{fig:partialps}) and the lower CH$_4$ abundance measured above Titan's haze, Titan-like conditions can be ruled out at better than 95\% confidence. Methane was also present on Earth during the Archean eon, which we model following previous calculations for TRAPPIST-1\,d \citep{meadows_feasibility_2023}. We find that the amounts of CH$_4$ in a 1 bar Archean Earth atmosphere are incompatible with our observations at the 3$\sigma$ level. This result holds even when considering the entire range of methane partial pressures from other models of Earth's atmosphere during this eon (approximately 10$^{-5}$ to 10$^{-2}$ bar; \citealp{catling_archean_2020}).

We similarly compare our results in the N$_2$-CO$_2$ and N$_2$-H$_2$O cases to the composition of modern Earth (as defined in \citealp{meadows_exoplanet_2017}, atmospheric water abundance on results from \citealp{kasai_h2o_2011,kelsey_atmospheric_2022}). We also show where a clear or cloudy Venus, a clear or cloudy early Mars or a thin-atmosphere modern Mars would lie in the N$_2$-CO$_2$ parameter space. The cloudy Venus-like composition assumes a 0.03 bar cloud deck in a thick CO$_2$-rich atmosphere. For early Mars, we adopt a 1.1 bar surface pressure including 1 bar of CO$_2$\citep{hu_nitrogen-rich_2022}. The cloudy early Mars case includes CO$_2$ ice clouds condensing out at $\sim 0.5$ bar, lowering the effective surface pressure \citep{turbet_global_2021}. For modern Mars, we adopt a 0.01 bar surface pressure.
We find that we can exclude very thick CO$_2$ atmospheres, such as a clear Venus composition, as well as the amounts of CO$_2$ expected for the ``Archean Earth'' 
\citep[similar in CO$_2$ content to the ``pre-industrial Earth'' scenario modeled by][]{meadows_feasibility_2023}. Although the CO$_2$ partial pressure in the atmosphere of Archean Earth could range from 10$^{-3}$ to 1 bar \citep{catling_archean_2020}, the value displayed in the Figure corresponds to the smallest realistic amounts of CO$_2$ in such a scenario and this model uncertainty does not impact our conclusions. 
Finally, we can rule out the ``early Mars'' scenario with better than 95\% confidence (even when accounting for the CO$_2$ ice cloud presence) -- although here also, the exact composition of Mars is uncertain and may have varied through the ages \citep{kite_geologic_2019}.
Only extremely thin or cloudy atmospheres, such as a 0.01-bar Mars-like atmosphere very rich in CO$_2$, or cloudy cases such to Venus' sulfuric acid clouds or Earth's water clouds remain marginally consistent with our observations (Figure \ref{fig:partialps}).

\begin{figure*}
    \centering
    \includegraphics[width=0.9\linewidth]{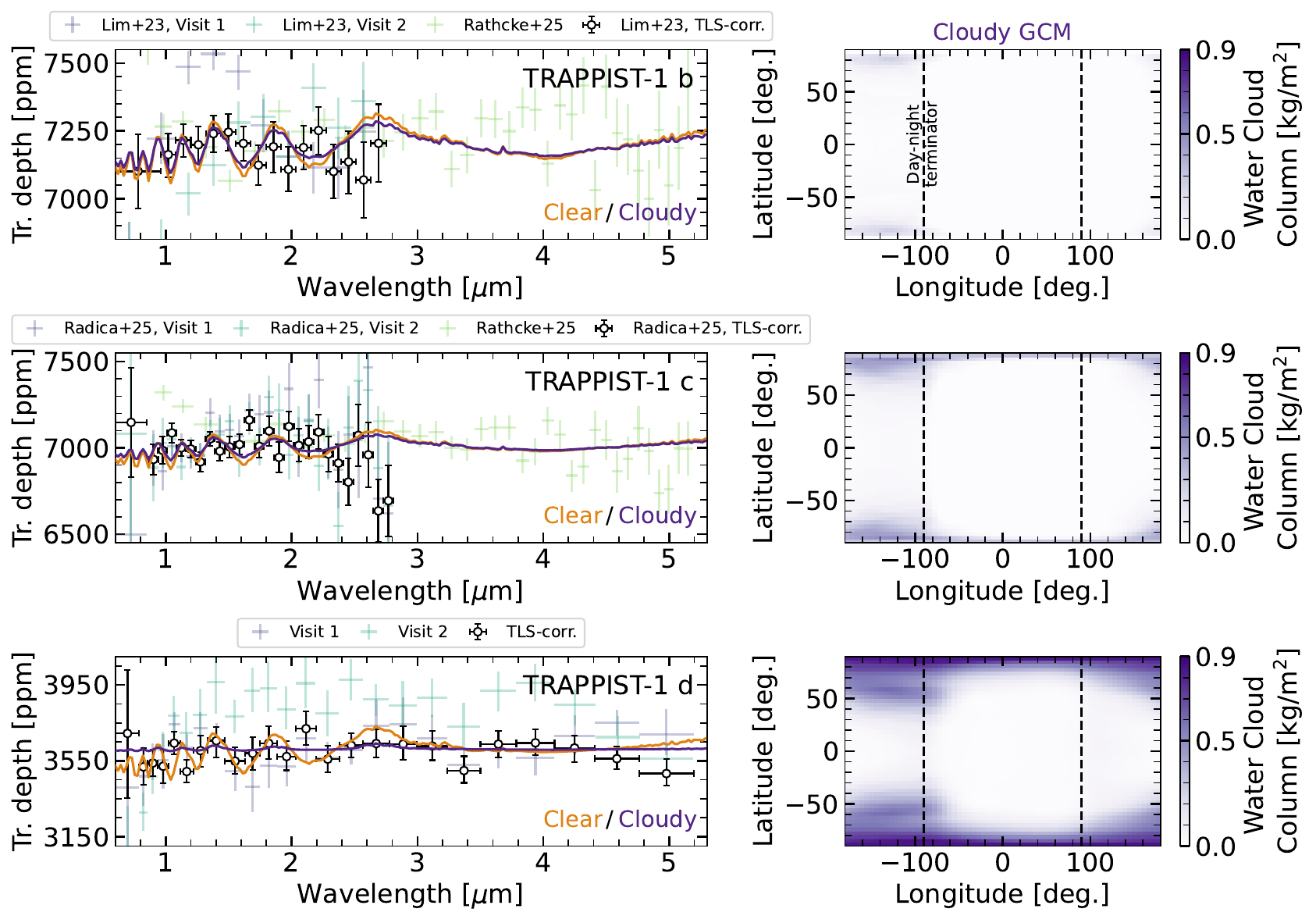}
    \vspace{5mm}
    \vspace{-5mm}\caption{Comparison of the transmission spectra of TRAPPIST-1 b, c, and d, with 3D global climate model (GCM) simulations. The left panels show the measured NIRISS/SOSS transmission spectra of TRAPPIST-1\,b and c \citep{lim_atmospheric_2023,RadicaT1c}, the NIRSpec/PRISM transmission spectra of TRAPPIST-1\,b and c \citep{rathcke_stellar_2025}, as well as the NIRSpec/PRISM transmission spectra of TRAPPIST-1\,d obtained in this work (colored points). Binned versions of the spectra for which \texttt{stctm} retrievals were published, after correction by the best-fitting TLS-only model (offset by 450 ppm for TRAPPIST-1 c) are shown as the black points. Binning was applied to the NIRISS/SOSS order 1 spectra of TRAPPIST-1\,b (five points together), to the TLS-corrected spectrum of TRAPPIST-1\,b (three points together), and to the NIRSpec/PRISM spectra of TRAPPIST-1\,b and c (10 points together), for visualization purposes. We show simulated spectra from 3D GCM simulations of a water-rich composition (10 bar H$_2$O + 1 bar N$_2$) on each planet, for cloudy (purple) and cloud-free (orange) conditions. While clouds barely affect the GCM model spectra in the cases of TRAPPIST-1\,b and c, they reach the terminator and flatten the transmission spectrum at the insolation of TRAPPIST-1\,d. The right panels illustrate this by showing the water cloud columns as a function of latitude and longitude for each cloudy GCM simulation (shades of purple, terminator shown with black dashed lines).}
    \label{fig:gcm}
\end{figure*}

\subsection{Implications of a ``hot start'' scenario on the detectability of a water-rich atmosphere}\label{sec:hot_start}



The lack of observed atmospheric absorption features in the transmission spectrum of TRAPPIST-1\,d (Figure \ref{fig:atmosphere_molbymol}) cannot conclusively rule out the presence of a high-mean molecular weight atmosphere, especially if the conditions are met for aerosols to form in the upper atmosphere at the terminator (Figure \ref{fig:met_psurf}). We turn to 3D global climate model (GCM) simulations in order to evaluate the possibility that TRAPPIST-1\,d's transmission spectrum can be explained by the formation of such aerosols in a water-rich atmosphere.

The traditional suite of GCMs used to define habitable conditions (i.e., liquid water at the planetary surface) for rocky planets in synchronous rotation around M dwarfs \citep{yang_stabilizing_2013,kopparapu_inner_2016,kopparapu_habitable_2017} assume ``cold start'' conditions (the water is initially condensed out) to delineate the inner edge of the habitable zone (HZ) as the lowest instellation required to trigger the loss of the water ocean through the runaway greenhouse effect. In such scenarios, the albedo feedback provided by low-altitude dayside clouds can be sufficient for TRAPPIST-1\,d to have habitable conditions. However, it is likely that planets instead start ``hot'' \citep{ikoma_water_2018,lichtenberg_super-earths_2024}: the gravitational potential energy released after formation, as well as the potential greenhouse effect of any accreted volatiles, can bring the interior of a planet to temperatures high enough to reach a completely molten state. The mantle of TRAPPIST-1\,d could then remain at least partially molten over geological timescales as it gradually crystallizes (e.g., \citealp{barr_interior_2018,grayver_interior_2022}).

In such a ``hot start'' scenario, we show that if abundant water is present in the atmosphere, cloud formation is predicted by GCM modeling. Further, TRAPPIST-1\,d's insolation places it in a regime where such clouds could essentially preclude the detection of water-rich atmospheres using transmission spectroscopy. Indeed, 3D GCM modeling predicts that the present-day orbit TRAPPIST-1\,d (currently receiving a flux of 1.125$S_\oplus$; Table \ref{table:pla_star_param}) would be interior to TRAPPIST-1's water condensation limit (at about 0.772$S_\oplus$), which is the limit in insolation flux where the planet would cool down enough over its lifetime to condense a surface liquid water ocean. Therefore, any water initially accreted by TRAPPIST-1\,d or outgassed by mantle geochemistry should remain in the atmosphere \citep{turbet_water_2023}. However, contrary to the inner planets TRAPPIST-1\,b and c, given its lower insolation, $\sim$mbar-level stratospheric water clouds form on the nightside of a water-rich N$_2$--H$_2$O atmosphere on TRAPPIST-1\,d and reach up to the terminator region (Figure \ref{fig:gcm}). Such high-altitude clouds would provide a plausible explanation for our observed flat transmission spectrum, given the current level of precision (Figure \ref{fig:gcm}). These results are seemingly contrasting with previous work that found that the presence of such stratospheric clouds in an Earth-like atmosphere does not impact atmospheric signatures in transmission \citep{doshi_stratospheric_2022}. However, the Earth-like stratospheric clouds are modeled at 600 to 100 mbar by \citet{doshi_stratospheric_2022}, which is 2 to 3 orders of magnitude deeper than the water clouds predicted to form on the TRAPPIST-1 planets \citep{turbet_water_2023}, highlighting the importance of full planet-specific GCM simulations for tidally locked rocky planets. In comparison, for CO$_2$-dominated atmospheres, the spectroscopic impact of CO$_2$ cloud condensation is very minimal for TRAPPIST-1\,d \citep{turbet_water_2023}. 

Finally, while H$_2$O clouds are stable across a wide variety of surface compositions, the atmosphere on TRAPPIST-1\,d could also be obscured by other cloud species expected at these temperatures. Specifically, nitrogen-bearing condensates such as ammonia (NH$_3$), ammonium chloride (NH$_4$Cl), or ammonium hydrosulfide (NH$_4$SH), as well as solid disulfide (S$_2$[s]) are expected for temperatures lower than $\sim400$\,K from crust-atmosphere equilibrium calculations \citep{herbort_atmospheres_2022}. A volcanically-active hot surface could support the condensation of graphite or soot-like particles (C[s]), while a thick atmosphere with a surface pressure of $\sim 100$ bar would allow for the condensation of solid sulfuric acid (H$_2$SO$_4$[s]) Venus-like clouds \citep{bullock_atmosphere_venus}. 



\section{Implications of complete atmosphere stripping on formation conditions}\label{sec:evolution_scenarios}

Considering the planet in its irradiation context, may cast further doubt as to the likelihood of even a cloudy H$_2$O-dominated atmosphere. In particular, 3D GCM simulations suggest that any water on TRAPPIST-1\,d would lie in its atmosphere in vapor (or condensed cloud) form, rather than in a liquid surface water ocean \citep{turbet_water_2023}. This distribution of the water inventory makes the planet more susceptible to water loss at its present-day orbital distance than the cooler habitable-zone planets TRAPPIST-1 e, f, and g, which likely exited their initial runaway greenhouse phase. The complete stripping of the atmosphere of TRAPPIST-1\,d is consistent with our observed transmission spectrum, and we consider the implications of such a bare-rock planet in terms of its formation conditions and of the likelihood of present-day atmospheres on the outer TRAPPIST-1 planets.

Numerous theoretical works have found that the TRAPPIST-1 planets are especially vulnerable to the loss of atmospheric volatiles due to the long pre-main-sequence period where the stellar XUV irradiation is particularly intense \citep{bolmont_water_2017,moore_role_2023,van_looveren_airy_2024}. This fragility against potential loss means that atmospheric retention is limited by each planet's initial volatile inventory, which would have to be large enough to prevent complete stripping or allow for significant atmospheric outgassing. 

Stripped atmospheres with surface pressures less than 1 bar are the most likely outcomes for TRAPPIST-1\,b and c, even for initial water inventories of up to $ \sim 100$ Earth oceans (close to the theoretical upper limit for the inner planets TRAPPIST-1 b, c, and d;  \citealp{krissansen-totton_predictions_2022,krissansen-totton_implications_2023}). However, if TRAPPIST-1\,d initially had a similar amount of volatiles to TRAPPIST-1\,b and c, its range of plausible present-day atmospheres, even under the impact of the same host star irradiation, could a-priori allow for 1--10 bar atmospheres \citep{krissansen-totton_predictions_2022}. The non-detection of such a thick atmosphere of TRAPPIST-1\,d would therefore suggest the inner planets of the TRAPPIST-1 system started out much dryer than initially believed. 

\begin{figure}
    \centering
    \includegraphics[width=0.9\linewidth]{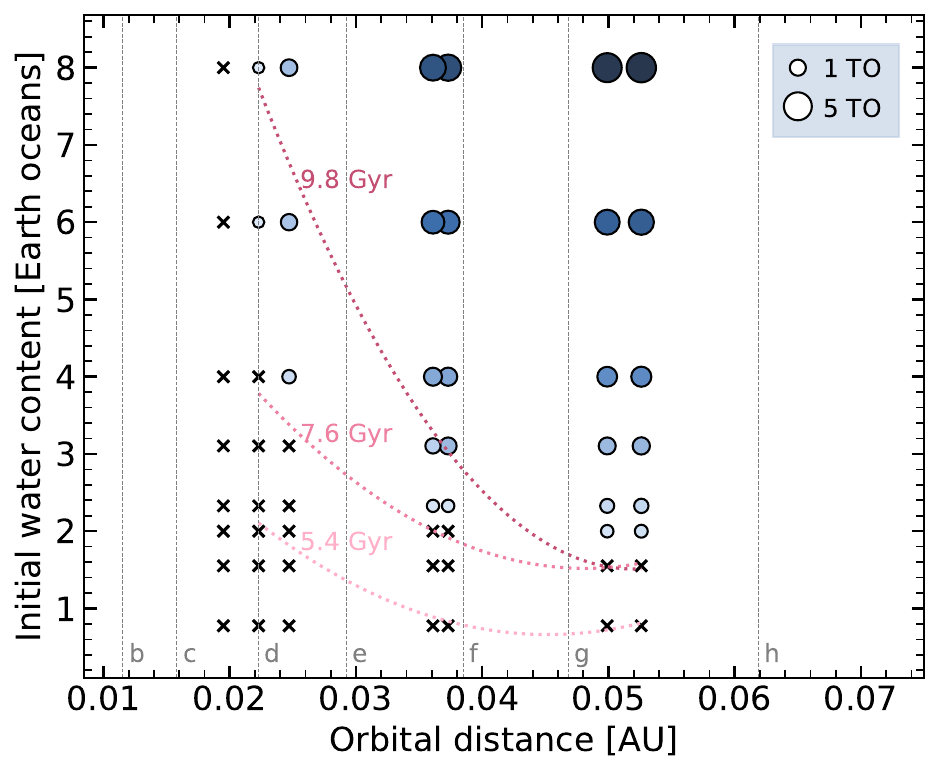}
    \vspace{5mm}
    \vspace{-5mm}\caption{Results from the planetary evolution models for TRAPPIST-1\,d as a function of initial water content and planetary orbital distance \citep[following][]{moore_water_2024}. The present-day orbital distances of the seven TRAPPIST-1 planets are indicated with grey dashed lines, and labeled at the bottom. Simulation results for planets evolved for 7.6 Gyr (median age of TRAPPIST-1) are represented by circles (size and color proportional to the final water content in Earth oceans; TO) or by `x' if the planet is completely desiccated at the end of the simulation. Pink lines show fits to the transition to desiccated planets when the simulation is run for 5.4, 7.6, and 9.8 Gyr. TRAPPIST-1\,d could have retained water in its interior and atmosphere over 5--10 Gyr if it migrated inwards only recently, or even in-situ if it initially accreted an initial water budget of more than $\sim 4$ Earth oceans.}
    \label{fig:water_retention}
\end{figure}

We use a box model prescription (see Section \ref{sec:water_loss_methods}; \citealp{moore_role_2023,moore_water_2024}) to estimate the vulnerability of various initial volatile inventories to stripping by stellar irradiation. 
TRAPPIST-1\,d, even at its present-day orbital distance, would likely have been shielded from complete water loss if it formed with more than about 4 Earth oceans of water (Figure \ref{fig:water_retention}). Such an initial inventory corresponds to a water mass fraction on the order of 0.2 to 0.3\% at formation, which is consistent with estimates of Earth's water mass fraction ranging from about 0.02 to 0.2\% (or 1 to 10 Earth oceans) depending on how much water is trapped by minerals in the mantle transition zone \citep{schmandt_dehydration_2014}.

Our evolution models therefore support a dry formation for the inner TRAPPIST-1 planets in the case of complete stripping of TRAPPIST-1\,d's atmosphere. In our simulations, TRAPPIST-1\,d retains a magma ocean over $\sim 500$\,Myr in our simulations in which water can be efficiently dissolved, essentially protecting the water inventory from escape throughout the very early and active stages of the star's evolution. However, our prescription for the water saturation limit of the magma ocean implies that if more water is initially present than can efficiently be dissolved in the mantle, atmospheric erosion can begin right from the start of the simulation. 
Alternatively, TRAPPIST-1\,d may have migrated inwards only recently, and could therefore have been shielded from desiccation by exiting the runaway greenhouse phase earlier in its evolution. 

Compared to TRAPPIST-1\,d, the outer TRAPPIST-1 planets are able to more effectively shield water from loss to space in surface liquid ocean reservoirs \citep{turbet_modeling_2018}. This conclusion is further supported by our evolution models run for TRAPPIST-1 e, f, and g. All three planets lie outside the runaway greenhouse limit and are, respectively, 1.8, 2.7, and 3.4 times more massive than TRAPPIST-1\,d: they would therefore have retained significant water inventories over $\sim 5$ to 10 Gyr provided they formed with at least a few Earth oceans of water (Figure \ref{fig:water_loss_efg}), even without invoking any migration. Further, the outer TRAPPIST-1 planets could have much larger initial volatile budgets than the inner three \citep{krissansen-totton_predictions_2022,krissansen-totton_implications_2023}, suggesting that even if TRAPPIST-1\,d is stripped, less-irradiated planets likely still bear atmospheres provided they accreted a moderate amount of water during or after their formation.








\section{Summary and Conclusions} \label{sec:conclusion}

We present the first 0.6--5.2$\mu$m \textit{JWST} transmission spectrum of the small temperate terrestrial planet TRAPPIST-1\,d, obtained from two transit observations with NIRSpec/PRISM. We detect small stellar flares in both visits and find that stellar contamination has an important impact on our observations, leading to 500-1,000~ppm signatures. Stellar contamination alone provides a satisfactory fit to our observations, without the need for any planetary atmosphere.  

The small flares we flag are apparent in the H$\alpha$ time series, while they are otherwise barely noticeable in the white light curves, stressing the importance of monitoring spectral lines affected by flares and the potential of even the lower resolution of NIRSpec/PRISM, compared to NIRISS/SOSS, for doing so. Although small, these flares have wavelength-dependent properties and require independent detrending for each spectroscopic light-curve to obtain robust transit spectra.

We find that stellar contamination is the dominant contributor to our transmission spectra, and that it varies from one visit to the next although they are only 4 days, or 1.2 stellar rotations apart.
We perform additional retrievals on the out-of-transit stellar spectrum, in order to compare the TLS-effect-inferred stellar surface heterogeneity distributions that impact joint TLS-atmosphere retrievals to prior expectations based on the baseline stellar spectrum. We introduce \texttt{exotune}, a sub-routine of the open-source \texttt{stctm} module, with capabilities to perform such retrievals, and suggest this method as a systematic cross-check as it provides complementary information to standard TLS retrievals. Out-of-transit spectral retrievals provide independent constraints on the photosphere temperature to be compared to inferences based on high-resolution spectra, which can be biased by e.g. incomplete line lists for M dwarfs \citep{jahandar_chemical_2024}. 
Improvements in model fidelity, for example using data-driven approaches \citep{Rackham_SAG21_2023,deWit_roadmap_2023} will be crucial to provide better stellar contamination corrections and to obtain even more precise limits on potential atmospheres on the TRAPPIST-1 planets.

Once the TLS effect is accounted for, our NIRSpec/PRISM observations reach a high sensitivity even to low scale height, high mean molecular weight atmospheres with absorption from H$_2$O, CO, CH$_4$, NH$_3$, CO$_2$ or SO$_2$. In particular, NIRSpec/PRISM's sensitivity to CO$_2$, which is out of the reach of NIRISS/SOSS, is essential because of its crucial role in dictating whether the conditions are met for surface liquid water on temperate rocky exoplanets \citep{fauchez_impact_2019}. The fact that the atmospheric signal contained in the transmission spectrum seems flat, or muted and without any haze-like slope enables us to place stringent constraints on the presence of an atmosphere on TRAPPIST-1\,d.

All thick or cloud-free (cloud-top pressures deeper than $0.01$ bar) hydrogen-rich ($\lesssim 100 \times$ solar metallicity) atmospheres are excluded by our observations at the $>3\sigma$ level. Since high-altitude aerosols are unlikely to form in hydrogen-rich temperate atmospheres \citep{morley_thermal_2015,moran_limits_2018}, we rather explore two main possibilities: either TRAPPIST-1\,d has an extremely thin, or cloudy, high mean molecular weight atmosphere, or it is airless.

For high mean molecular weight scenarios, we demonstrate that the precision reached by our observations can rule out compositions similar to solar system terrestrials: a cloud-free Titan composition, or even compositions analogous to Archean Earth, early Mars, or a cloud-free Venus are ruled out by our observations with greater than 95\% confidence (2$\sigma$). Meanwhile, our spectra remain marginally consistent within uncertainties (near the 2$\sigma$ level) with an extremely thin Mars-like CO$_2$ atmosphere, a Venus-like scenario with high-altitude clouds, or the CO$_2$ signature of a modern Earth composition when accounting for Earth's water clouds. 

Our observations cannot yet completely exclude other potential atmosphere scenarios for TRAPPIST-1\,d which were predicted in the literature: for instance, 3D GCM modeling demonstrates that a water-rich atmosphere may form high-altitude $\sim$mbar water clouds that reach the terminator, obscuring any features in transmission \citep{fauchez_impact_2019,turbet_water_2023}. This issue is likely to impact atmosphere detectability in transmission for most habitable-zone terrestrial planets \citep{komacek_clouds_2020}. As an alternative to transmission spectroscopy, reconnaissance eclipse photometry using JWST/MIRI's 15$\mu$m bandpass could confirm the $\sim 200$ppm most optimistic airless scenario (i.e. no heat redistribution, zero-albedo surface) at 3$\sigma$ confidence with 7 eclipse observations \citep{doyon_temperate_2024}. Further modeling of the 3D impact of the condensation of water, as well as carbon-, nitrogen-, and sulfur-bearing species will be crucial for the interpretation of future transmission and emission studies of terrestrial exoplanets.

If TRAPPIST-1\,d is airless, our evolution models predict that TRAPPIST-1 b, c, and d must have formed relatively dry, with an initial water inventory of less than about 4 Earth oceans. Because of its higher likelihood of atmospheric retention, atmospheric reconnaissance of TRAPPIST-1\,d provides more leverage on the inner planets' initial volatile inventories and the outer planets' likelihood of atmosphere presence than TRAPPIST-1\,b and c (vulnerable to complete atmosphere loss even if they initially accreted about 100 Earth oceans; \citealp{krissansen-totton_implications_2023}). We find that even complete atmosphere loss for TRAPPIST-1\,d would not preclude atmosphere presence for the outer HZ planets TRAPPIST-1 e, f, and g, as, contrary to the inner planets, they could retain water thanks to efficient shielding in the interior and in the condensed surface liquid water reservoir even if they initially accreted only a few Earth oceans in volatiles.







~

~
We thank the anonymous referee for comments that improved the quality of this manuscript. C.P.-G. also thanks D. Samra and J. Bean for helpful discussions. This work is based on observations with the NASA/ESA/CSA James Webb Space Telescope, obtained at the Space Telescope Science Institute (STScI) operated by AURA, Inc. This project was undertaken with the financial support of the Canadian Space Agency. All of the data presented in this paper were obtained from the Mikulski Archive for Space Telescopes (MAST) at the Space Telescope Science Institute. The data used in this paper can be found in MAST at the following DOI: \dataset[10.17909/2jzw-7m72]{http://dx.doi.org/10.17909/2jzw-7m72}. C.P.-G. acknowledges support from the NSERC Vanier scholarship, and the Trottier Family Foundation. C.P.-G. also acknowledges support from the E. Margaret Burbidge Prize Postdoctoral Fellowship from the Brinson Foundation. M.R.\ acknowledges funding from NSERC, Fonds de recherche du Qu\'{e}bec -- Nature et technologies (FRQNT) and iREx. R.A. acknowledges the SNSF support under the Post-Doc Mobility grant P500PT\_222212 and the support of iREx. M.T. acknowledges support from the Tremplin 2022 program of the Faculty of Science and Engineering of Sorbonne University. M.T. acknowledges support from BELSPO BRAIN (B2/212/PI/PORTAL). D.B., K.M. and M.K. acknowledge support from InitiaSciences scholarships within the InitiaSciences research mentoring program. O.L. acknowledges financial support from the FRQNT. A.L’H. acknowledges support from the FRQNT under file \#349961. L.D. is a Banting and Trottier Postdoctoral Fellow and acknowledges support from the Natural Sciences and Engineering Research Council (NSERC) and the Trottier Family Foundation. J.T. acknowledges funding support by the TESS Guest Investigator Program G06165. This project has been carried out within the framework of the National Centre of Competence in Research PlanetS supported by the Swiss National Science Foundation under grant 51NF40\_205606. S.P.\ acknowledges the financial support of the SNSF. D.J.\ is supported by NRC Canada and by an NSERC Discovery Grant.


\software{\texttt{Eureka!},
          \texttt{jwst}\footnote{\url{https://github.com/spacetelescope/jwst}},
          \texttt{emcee}\footnote{\url{https://github.com/dfm/emcee})} \citep{foreman-mackey_emcee_2013},
          \texttt{corner}\footnote{\url{https://github.com/dfm/corner.py}} \citep{foreman-mackey_cornerpy_2016},
          \texttt{batman}\footnote{\url{https://github.com/lkreidberg/batman}} \citep{kreidberg_batman_2015},
          PHOENIX \citep{husser_new_2013},
          \texttt{MSG}\footnote{\url{https://github.com/rhdtownsend/msg}} \citep{townsend_msg_2023},
          \texttt{astropy}\footnote{\url{https://www.astropy.org/}} \citep{astropy_collaboration_astropy_2013},
          \texttt{numpy}\footnote{\url{https://github.com/numpy/numpy}} \citep{harris_array_2020}, 
          \texttt{stctm} \footnote{\url{https://github.com/cpiaulet/stctm} } (\citealp{stctm_zenodo_temp}, Piaulet-Ghorayeb, subm.),          \texttt{matplotlib}\footnote{\url{https://github.com/matplotlib/matplotlib}} \citep{hunter_matplotlib_2007}, \texttt{scipy}\footnote{\url{https://github.com/scipy/scipy}} \citep{virtanen_scipy_2020}
          },
          \texttt{FastChem}\footnote{\url{https://github.com/NewStrangeWorlds/FastChem}} \citep{stock_fastchem:_2018}







\appendix
\setcounter{figure}{0}
\renewcommand{\thefigure}{A\arabic{figure}}
\setcounter{table}{0}
\renewcommand{\thetable}{A\arabic{table}}

\section{Stellar and planetary parameters}
We present the key stellar and planetary parameters used in this work and inferred from our analysis in Table \ref{table:pla_star_param}.

\begin{table*}
\caption{Stellar and planetary parameters adopted or refined in this work. We combine the constraints on the planet's radius and orbital parameters from both NIRSpec/PRISM visits to obtain the values quoted here. The derived values are obtained using Monte Carlo sampling. The quoted orbital parameters from \citet{agol_refining_2021} correspond to the osculating Jacobi elements inferred from a transit timing analysis at the start time: BJD$_\mathrm{TDB}$ -2,450,000 = 7257.93115525 days, and derived parameters from a photodynamical analysis. Since the planet's orbit evolves over time, the quoted values of the orbital parameters obtained from our fit to the two NIRSpec/PRISM transits can be considered to be a measurement of the planet's orbital separation and inclination at the time of the observations.}
\centering
\begin{tabular}{lcc}
\hline
\hline
Parameter      & Value             & Source         \\

\hline
\hline
TRAPPIST-1 (star)&  & \\
\hline
Spectral type & M8.0V & \citet{liebert_ri_2006, gillon_seven_2017} \\
Effective temperature $T_\mathrm{eff}$ [K] & 2566 $\pm$ 26 & \citet{agol_refining_2021} \\
Metallicity [Fe/H]& 0.052$\pm$0.073 & \citet{davoudi2024} \\
Age [Gyr]& 7.6$\pm$2.2 & \citet{burgasser_age_2017} \\
Stellar mass $M_\star$ [$M_\odot$]& 0.0898$\pm$0.0023 & \citet{mann_how_2019} \\
Stellar radius $R_\star$ [$R_\odot$]& 0.1192$\pm$0.0013 & \citet{agol_refining_2021} \\
log$_{10}$ $\left(g_\star \left[cm/s^2\right]\right)$ & $5.2396^{+0.0056}_{-0.0073}$ & \citet{agol_refining_2021} \\
Stellar luminosity $L_\star$ [$L_\odot$]& 0.000553$\pm$0.000019 & \citet{agol_refining_2021} \\
Stellar rotation period [day]& 3.30$\pm$0.14 & \citet{luger_seven-planet_2017} 
\\
\hline
\hline
TRAPPIST-1\,d (planet)&  & \\
\hline
Planetary radius $R_p$ [$R_\oplus$] & $ 0.804^{+ 0.014}_{- 0.012}$  & This paper \\
Planetary mass $M_p$ [$M_\oplus$] & 0.388$\pm$0.012 & \citet{agol_refining_2021} \\
Orbital period $P$ [day] & 4.049219$\pm$0.000026 & \citet{agol_refining_2021} \\
Scaled semi-major axis $a/R_\star$ & $40.15 \pm 0.18$ & This paper \\
Orbital separation $a$ [AU]& $0.0223 \pm 0.0003$& Derived, this paper\\
Impact parameter $b$ & $0.06 \pm 0.05$& This paper \\
Inclination $i$ [deg] & $89.909 ^{+ 0.068}_{ - 0.071}$ & Derived, this paper \\
Insolation $S_\mathrm{p}$ [$S_\oplus$]& $1.119 ^{+ 0.047 }_{- 0.044 }$ & Derived, this paper \\
Equilibrium temperature &  & \\
~~~~- $T_\mathrm{eq, A=0.0}$ [K]& $ 286 \pm 3 $ & Derived, this paper \\
~~~~- $T_\mathrm{eq, A=0.3}$ [K]& $ 262 \pm 3$ & Derived, this paper \\

\hline
\label{table:pla_star_param}
\end{tabular}
\end{table*}

\setcounter{figure}{0}
\renewcommand{\thefigure}{B\arabic{figure}}
\setcounter{table}{0}
\renewcommand{\thetable}{B\arabic{table}}

\section{Light-curve analysis methodology}

\subsection{Broadband light-curve fitting}\label{sec:broadband_lc}
\begin{figure}
    \centering
    \includegraphics[width=0.95\textwidth]{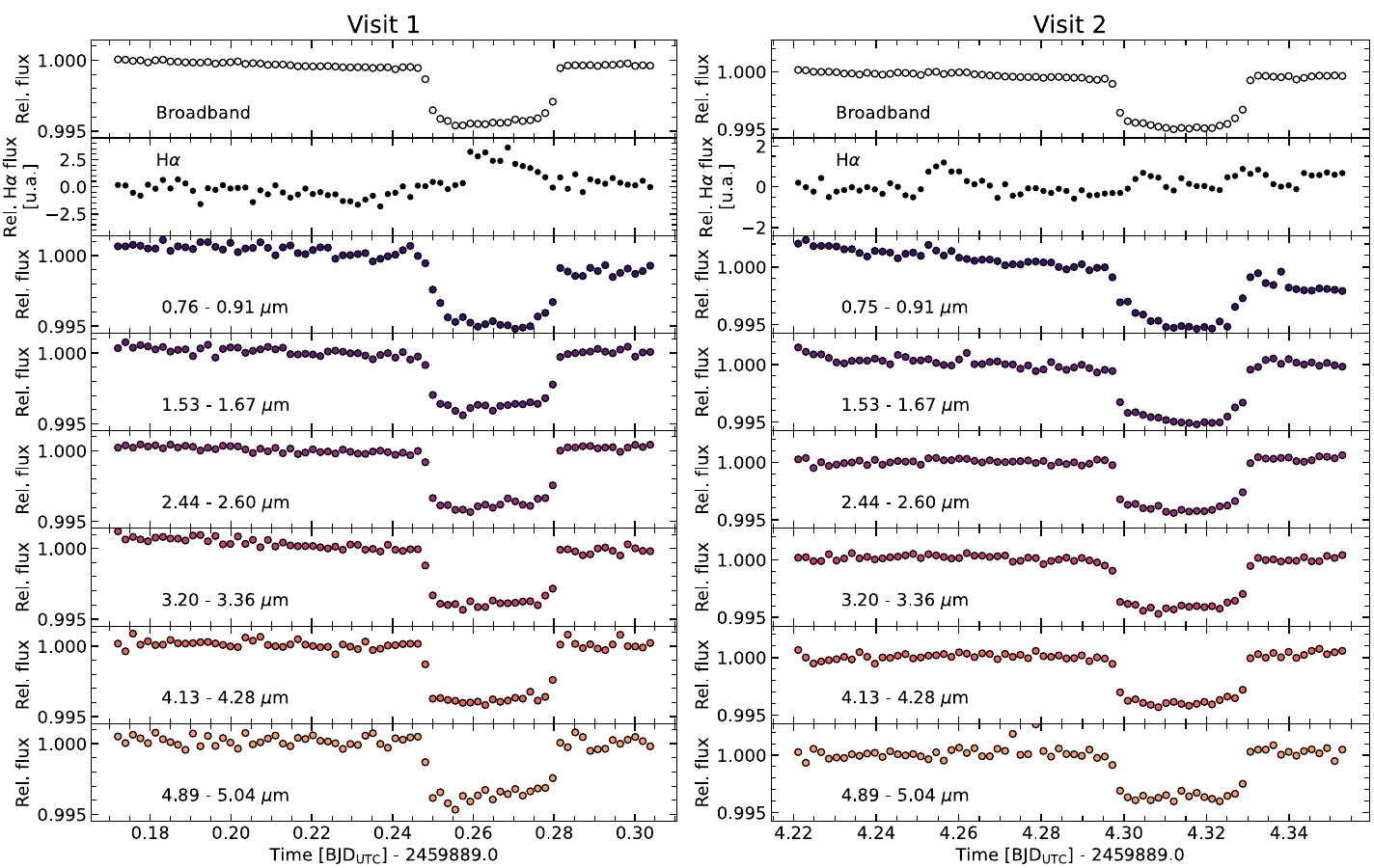}
    \vspace{5mm}
    \vspace{-5mm}
    \caption{For visit 1 (left) and visit 2 (right), we show the raw broadband light curves (top panels), along with the H$\alpha$ flux (in arbitrary units, scaled by $\times 10^{4}$; second panels) and a few example raw spectroscopic light curves (bottom panels, with wavelength bins labeled in the bottom left corners). For clarity, we bin 100 data points together to obtain the curves shown in this figure. The spectroscopic light curves exhibit different slopes (generally more downwards for shorter wavelengths) and wavelength-dependent features correlating with the small flares flagged in the H$\alpha$ time series.}
    \label{fig:raw_lightcurves}
\end{figure}

\begin{figure}
    \centering
    \includegraphics[width=0.45\textwidth]{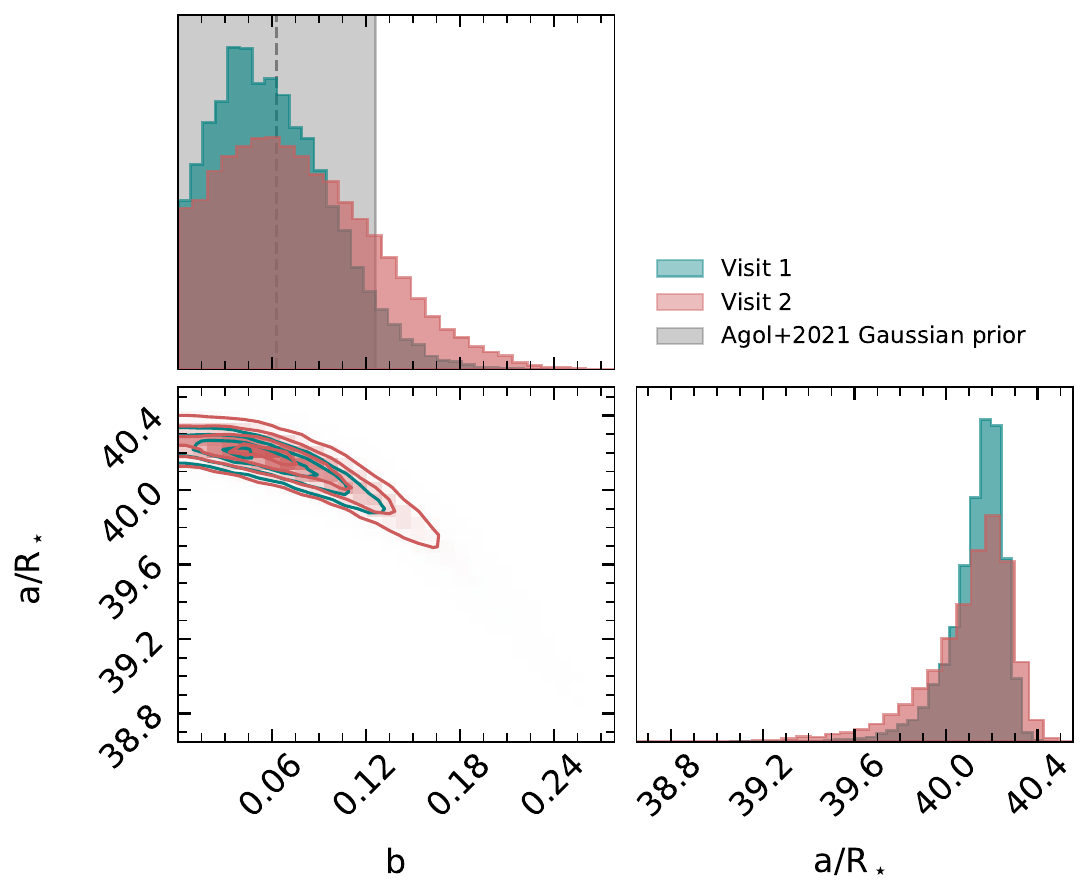}
    \vspace{5mm}
    \vspace{-5mm}
    \caption{Joint (bottom left) and marginalized (top left, and bottom right) posterior distributions on the impact parameter $b$ and the scaled semi-major axis $a/R_\star$ from the white light-curve fits to both visits (colored joint and marginalized distributions). The median and 1$\sigma$ uncertainty region corresponding to the Gaussian prior on the impact parameter are indicated by the gray-shaded region.}
    \label{fig:b_aRs}
\end{figure}

We construct our white light curves for each visit by summing all spectroscopic light curves, after discarding all wavelength channels where pixels are flagged as saturated in the median-in-time image. We find that this reduces the point-to-point scatter in the broadband light-curve by $\approx$ 15 to 18\% in both visits. Raw white and spectroscopic light curves are shown in Figure \ref{fig:raw_lightcurves}. We then perform individual broadband light-curve fits to each light-curve using the \texttt{ExoTEP} pipeline \citep{benneke_spitzer_2017,benneke_sub-neptune_2019}.

We find that the light curves of both visits exhibit correlated noise on top of the transit signal. Although the light-curve variations do not track with instrumental metrics such as the position or width of the trace in the spatial or dispersion direction, they correlated well with the recorded variations in the integrated flux around the H$\alpha$ line (Figure \ref{fig:wlc_Halpha}). As with previous studies of TRAPPIST-1 \citep{lim_atmospheric_2023,howard_characterizing_2023,RadicaT1c}, we therefore attribute the variations we observe to stellar flares, although in our case the flares have very small amplitudes in terms of their imprint on the white light curve. They still impact our analysis given their unfortunate timing in and near the transit, suggesting that even small flare events can be problematic when extracting small-planet transmission spectra around late-type M dwarfs. 

Direct detrending of the light curves against the H$\alpha$ flux time series itself was unsuccessful, as we observe flare-dependent time delays between the appearance of the flare in H$\alpha$ and the corresponding rise in flux observed in the white light curves. This is not surprising, given that flares have time lags between the release of energy at different wavelengths \citep{howard_characterizing_2023}. We therefore use a Gaussian Process to model this astrophysical noise using a non-parametric model.

We test two GP kernels: the Mat\'{e}rn 3/2 kernel, and a simple harmonic oscillator (SHO) kernel with the $Q$ parameter fixed to $1/\sqrt{2}$ (critically damped SHO term, as in e.g., \citealp{radica_muted_2024}).
The Mat\'{e}rn kernel is a stationary kernel that generalizes the squared-exponential kernel with a parameter $\nu$ which controls the smoothness of the resulting function. While $\nu=1/2$ corresponds exactly to the squared-exponential kernel, $\nu=3/2$ is the variant that produces once differentiable functions and is implemented in \texttt{celerite} \citep{celerite_2017}. The critically-damped SHO kernel, on the other hand, is more appropriate to model granulation noise than one-off flare events \citep{coulombe2025highlyreflectivewhiteclouds}, but we test it because of the small amplitude of the flares we model, and for comparison with other terrestrial M dwarf planet light curves where it was previously applied \citep{cadieux_transmission_2024,RadicaT1c}. 

Our full systematics model combines a linear slope with time, frequently observed in other JWST/NIRSpec datasets (e.g., \citealt{rustamkulov_jwst_2023}), the GP parameters, and an extra jitter term added in quadrature to the diagonal of the covariance matrix. For each GP kernel, we vary the correlation amplitude parameter, as well as the correlation timescale in the fit. 

In our astrophysical model, we fit the planet-to-star radius ratio, the time of transit center, the planet's impact parameter, and its scaled semi-major axis (Table \ref{table:wlc_param}). After testing limb-darkening parameterizations where we fit a quadratic law with either the $(q_1, q_2)$ triangular sampling basis \citep{kipping_efficient_2013} or the $(u_1, u_2)$ uniform basis (see following subsection for the details), we adopt as fixed limb-darkening coefficients the best-fitting values obtained from the fit to the first NIRSpec/PRISM transit of TRAPPIST-1\,g (Benneke et al., submitted), which was not affected by flares. As in \citet{lim_atmospheric_2023}, we use a wide Gaussian prior on the impact parameter of $\mathcal{N}(0.063,0.063)$ informed by previous photodynamical modeling of the TRAPPIST-1 planets \citep{agol_refining_2021}: even with our good time sampling of the light curve, visit 2 in particular suffers from small flares near ingress and egress that could alter the inferred transit duration (Figure \ref{fig:b_aRs}). We adopt wide uninformative priors on all the other parameters. Our final model combines the astrophysical (transit) component, and the systematics model described above. 
This full model provides a satisfactory fit to both light curves (Figure \ref{fig:wlc_Halpha}), with Gaussian-distributed residuals. We obtain consistent fit quality and inferred planetary parameters regardless of the GP kernel used, but we choose to adopt the results obtained with the Matérn 3/2 GP, since its covariance structure is more appropriate to model the one-off flare events we observed. A more detailed exploration of the impact of GP model assumptions is offered in the next subsection. 

\begin{table*}
\caption{Fitted planetary parameters from the \texttt{ExoTEP} white light-curve fits of both \textit{JWST} NIRSpec/PRISM transits of TRAPPIST-1\,d. The planet-to-star radius ratio, scaled semi-major axis, and impact parameter are consistent within their uncertainties for both visits. The impact parameter was fitted with a prior of $\mathcal{N}(0.063,0.063)$ obtained by \citet{agol_refining_2021}.}
\centering
\begin{tabular}{lccc}
\hline
\hline
Fitted parameters      & Visit 1 & Visit 2         \\
\hline

Mid-transit time $T_c$ [BJD$_\mathrm{TDB}$]     &  $2459889.26447 \pm 0.00002$ & $2459893.31397 \pm 0.00003$  \\
Planet-to-star radius ratio $R_p/R_*$ & $0.0614 \pm 0.0005$  & $0.0624_{-0.0006}^{+0.0007}$      \\
Semi-major axis $a/R_*$   &   $40.16_{-0.14}^{+0.08}$ & $40.14_{-0.22}^{+0.12}$\\
Impact parameter $b$      &   $0.06_{-0.03}^{+0.04}$ & $0.07_{-0.05}^{+0.06}$ \\

\hline

\label{table:wlc_param}
\end{tabular}
\end{table*}

\subsection{Spectroscopic light-curve fitting}\label{sec:spec_lc}


We obtain the transmission spectrum of TRAPPIST-1\,d for each visit by fitting the spectroscopic light curves using \texttt{ExoTEP}, and explore the impact of our choice of limb-darkening and GP kernel prescription on the final spectrum. We find that flares impact the determination of limb-darkening parameters in the second visit, and that the correlated noise we observe has chromaticity that needs to be accounted for. 

In all the fits described below, we follow a similar procedure to the white light-curve analysis and jointly fit the astrophysical transit model described by the planet-to-star radius ratio $R_p/R_\star$, and a systematics model including parameters that capture instrumental and astrophysical noise, to each spectroscopic light curve. We fix the orbital parameters ($b$, $a/R_\star$), and the mid-transit time $T_c$ to the best-fit parameters from the white light-curve fit.

Our first step consists in obtaining a first-pass spectrum at the full instrumental resolution, with one light-curve extracted for each column along the dispersion direction. The goal of this step is to identify any outlier columns that should be discarded for the final analysis, especially in the region of the detector where ramps were affected by saturation. For this spectrum, we adopt \texttt{ExoTiC-LD} \citep{Grant2024,husser_new_2013} 1D limb-darkening coefficients, as our light curves do not have the signal-to-noise required to fit them. We choose this approach despite the fact that stellar model fidelity can be an issue for late M dwarfs since our aim is to identify columns that are outliers relative to their neighbors rather than to extract a valid spectrum for further analyses. We fit a simple transit model with a slope as the systematics model. We find that the following outlier columns should be discarded, as they were $>4\sigma$ outliers to the neighboring pixels: for visit 1, the column centered at 1.4115 $\mu$m, for visit 2, the column centered at 1.0663 $\mu$m, and for both visits, the three columns spanning 1.5716 to 1.6117 $\mu$m. To obtain the final spectra, we therefore construct the binned light curves by ignoring these four columns in each visit.

We perform sensitivity analyses to evaluate the impact of the choice of the limb-darkening parameters and of the correlated noise model on our transmission spectrum. We use as a benchmark for the limb-darkening coefficients the values that were fitted to the first NIRSpec/PRISM transit of TRAPPIST-1\,g (Benneke et al., submitted), as this observation was not affected by small or large flares that could have a wavelength-dependent impact on the light curves and the determination of the star's intensity profile (and is largely free of unocculted stellar surface heterogeneity signatures). In order to leverage these limb-darkening coefficients, we bin the light curves following the wavelength bins of Benneke et al. (submitted) prior to the spectroscopic fit.

\subsubsection{Limb-darkening prescription}\label{sec:LD}

\begin{figure}
    \centering
    \includegraphics[width=0.85\textwidth]{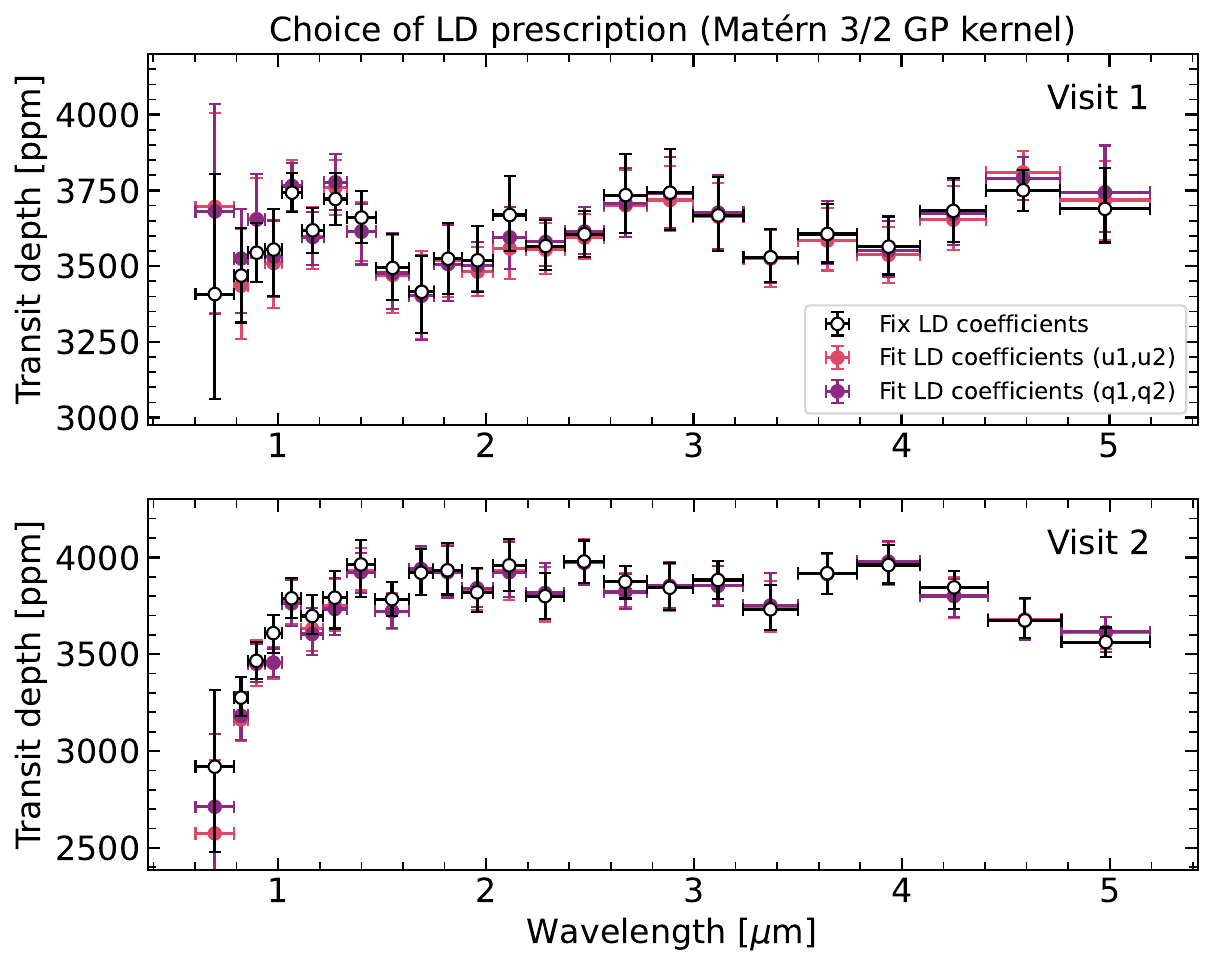}
    \vspace{5mm}
    \vspace{-5mm}\caption{Assessment of the impact of either fixing the limb-darkening parameters to the best-fit values from the first NIRSpec/PRISM transit of TRAPPIST-1\,g (Benneke et al., submitted; shown in black) or fitting them in the broadband and spectroscopic light-curve fits using the $[u_1,u_2]$ basis (pink) or the $[q_1,q_2]$ parameterization (magenta). The top (bottom) panel shows the results for the first (second) visit. Our spectra are largely insensitive to the choice of limb-darkening parameterization, except at short wavelengths in both visits (more pronounced for visit 2). The presence of small stellar flares during the first transit and near the ingress and egress of the second transit likely alter the determination of limb-darkening parameters from the light curves at short wavelengths.}
    \label{fig:test_LD}
\end{figure}

We evaluate the impact of fitting, or fixing the limb-darkening coefficients to independently-determined values on the final transmission spectrum. For this series of tests, the systematics model consists in a slope with time, a Matérn 3/2 GP where the GP amplitude and timescale are fitted for each wavelength bin (an assumption that is explored further in the next subsection), and an additional white noise scatter term. The astrophysical transit model describing each of the spectroscopic light curves has either 1 or 3 fitted parameters: $R_p/R_\star$, and the two coefficients of a quadratic limb-darkening law if they are fitted. 

We perform three tests: (1) holding the limb-darkening parameters fixed to the best-fit values fitted for TRAPPIST-1 in each of the same wavelength bins from Benneke et al. (submitted); (2) fitting the quadratic limb-darkening parameters $q_1$ and $q_2$ \citep{kipping_efficient_2013} in each bin; (3) fitting instead $u_1$ and $u_2$ in order to alleviate a potential bias from the $(q_1,q_2)$ prior on the transmission spectrum \citep{coulombe_biases_2024}. 

We find excellent agreement between the three versions of the transmission spectrum for each visit (Figure \ref{fig:test_LD}), which suggests that at least at wavelengths longer than $\sim 1.5 \mu$m, we can reliably constrain the limb-darkening parameters from our spectrum. However, we note a discrepancy at the shorter-wavelength end $\lambda \lesssim 1.5 \mu$m, resulting in systematically-lower inferred transit depths in the second visit when the limb-darkening parameters are fitted, compared to when they are fixed. Such short-wavelength differences may be driven by flares that affect the spectrum more strongly at shorter wavelengths \citep{howard_characterizing_2023} and are observed coinciding with the timing of ingress and egress for the second TRAPPIST-1\,d visit. Since this difference with the fixed-limb darkening version is not observed for the first visit, while a real offset would be systematic for a shared host star, we elect to use the spectrum obtained with the limb-darkening coefficients fitted to the transit of TRAPPIST-1 g.

\subsubsection{GP parameterization}\label{sec:GP}

\begin{figure}
    \centering
    \includegraphics[width=0.85\textwidth]{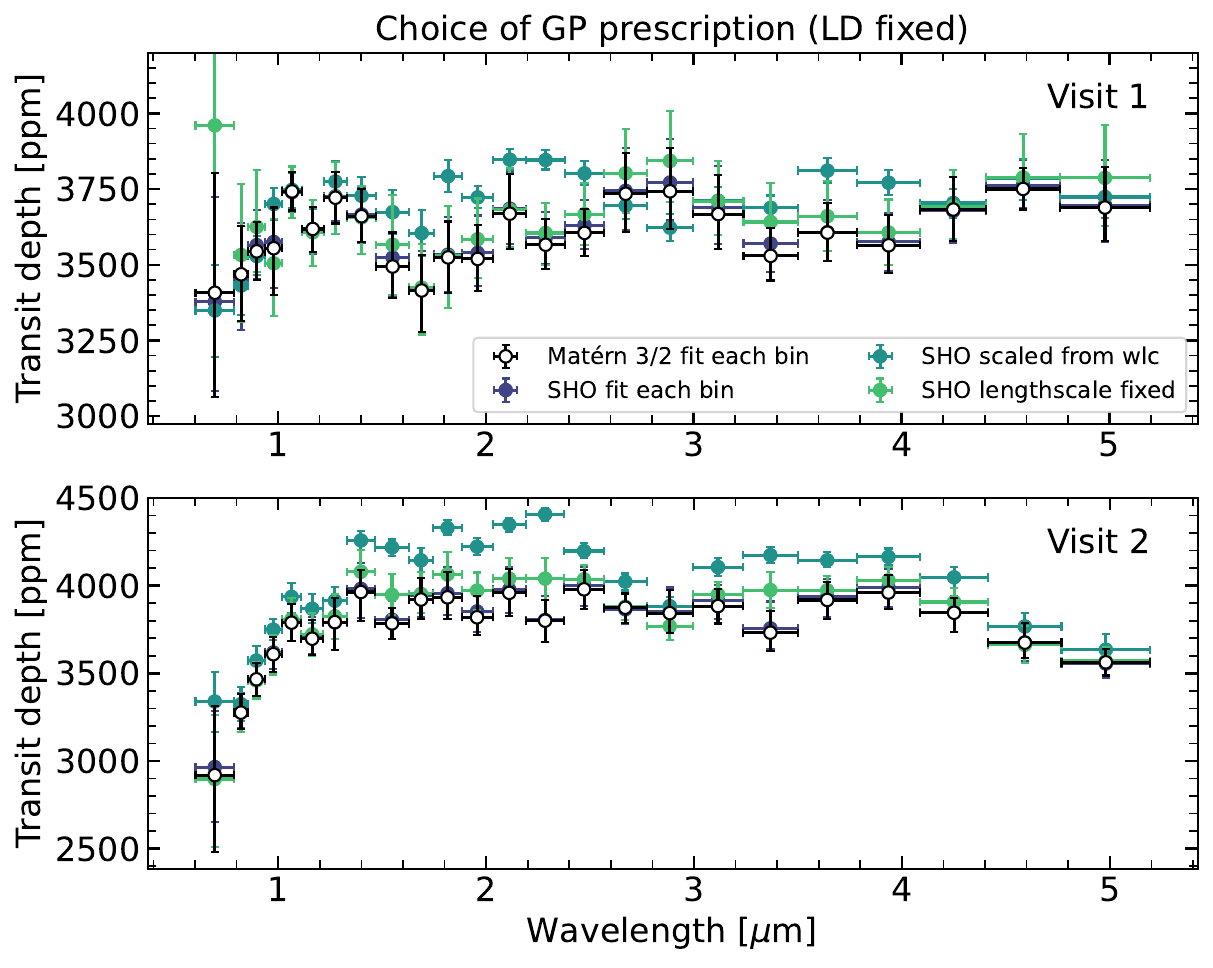}
    \vspace{5mm}
    \vspace{-5mm}\caption{Assessment of the impact of the GP parameterization adopted for spectroscopic fits, for visit 1 (top) and visit 2 (bottom), in the case where the limb-darkening parameters are kept fixed to prescribed values (see Figure \ref{fig:test_LD}). We tested two different GP kernels: the Matérn 3/2 kernel (black), and the simple harmonic oscillator (SHO) kernel (shades of blue). When the GP amplitude and lengthscale are fitted in each wavelength bin (black and dark blue points), the final spectrum is virtually identical. Assuming that the GP kernel lengthscale that corresponds to the best-fitting white light-curve model is also shared across all wavelength bins yields a consistent spectrum to a case where it is fitted, but the light-curve fit quality from 1.5 to 4$\mu$m is improved by fitting independent values (light green). Finally, the best-fitting GP component from the white-light-curve fit does not match well the spectroscopic light curves over most of the wavelength range (dark green), and leads to underestimated error bars because of the assumed correlated noise structure.}
    \label{fig:compare_GP}
\end{figure}

The treatment of correlated noise at the spectroscopic light-curve fit stage relies on our assumptions about the chromaticity and functional approximation of the flare signature. We test several scenarios: 
\begin{enumerate}
    \item \textit{Achromatic signature:} Scaling the mean GP model extracted from the best-fit to the white light-curve for each light curve: maintaining a mean of zero for the systematics model, but fitting a multiplicative parameter that allows to vary the amplitude of the variations relative to zero residuals.
    \item \textit{Achromatic correlation timescale:} Fitting the amplitude of the GP to each spectroscopic light-curve individually, but keeping the GP timescale fixed to the best-fit value from the white light-curve fit. In this scenario, contrary to the first one, the correlation structure of the residuals is inferred from each spectroscopic light-curve (using the position of neighboring points) rather than by the white light-curve residuals.
    \item \textit{Chromatic timescale and amplitude:} Fitting both the GP amplitude and timescale to each spectroscopic bin. For this scenario, we tested the impact of switching the GP kernel between the Matérn 3/2 and SHO kernel.
\end{enumerate}
For each of these tests, we hold the limb-darkening parameters fixed to the Benneke et al. (subm.) values.

We find that, while approaches (2) and (3) provide consistent spectra, method (1) yields a very different spectrum (Figure \ref{fig:compare_GP}). Therefore, we infer that the \textit{timescale} over which the flare affects the light curves is largely wavelength independent, but that the exact timing and shape of the flare signal is chromatic (Figure \ref{fig:raw_lightcurves}). The choice of GP kernel does not affect the spectrum, and we choose to use the Matérn 3/2 GP result in the rest of the analysis as it is inherently tailored to match the type of non-periodic flare signal we are correcting. Finally, we find that when fitting the GP timescale in each wavelength bin, it can be satisfactorily constrained by each individual light curve. Since this approach better captures the uncertainty on the GP timescale rather than assuming one value extracted form the white light curve, we move forward with the spectrum obtained with method (3) (see fitted spectroscopic light curves in Fig. \ref{fig:wlc_Halpha}). The transmission spectra we obtain are provided in Table \ref{tab:merged_transm_spec}.

\begin{table}
\begin{tabular}{cccccccc}
\hline
\hline
  & \multicolumn{3}{c}{Visit 1} & \multicolumn{3}{c}{Visit 2}  \\
Wavelength $[\mu$m] & Depth [ppm] & +1$\sigma$ [ppm] & -1$\sigma$ [ppm] & Depth [ppm] & +1$\sigma$ [ppm] & -1$\sigma$ [ppm] \\

\hline
   0.60 -- 0.79    &  3407.6     & 396.4    &  345.7 &  2919.2     & 394.9    &  438.2 \\
   0.79 -- 0.86    &  3468.6     & 158.5    &  155.0 &  3275.7     & 108.2    &  95.6 \\
   0.86 -- 0.94    &  3544.3     & 96.9    &  96.0 &  3465.8     & 95.4    &  93.3 \\
   0.94 -- 1.02    &  3554.6     & 134.6    &  153.1 &  3608.7     & 95.7    &  101.7 \\
   1.02 -- 1.11    &  3742.3     & 64.3    &  63.1 &  3788.6     & 106.1    &  102.3 \\
   1.11 -- 1.22    &  3618.3     & 73.4    &  76.1 &  3697.0     & 107.6    &  93.1 \\
   1.22 -- 1.33    &  3721.4     & 86.0    &  85.1 &  3791.7     & 139.9    &  158.9 \\
   1.33 -- 1.47    &  3660.3     & 88.7    &  84.4 &  3963.5     & 125.8    &  166.4 \\
   1.47 -- 1.63    &  3494.0     & 112.5    &  104.5 &  3784.4     & 89.6    &  88.6 \\
   1.63 -- 1.75    &  3415.0     & 117.2    &  137.0 &  3920.4     & 122.5    &  112.7 \\
   1.75 -- 1.89    &  3524.4     & 118.6    &  117.6 &  3933.1     & 141.9    &  124.7 \\
   1.89 -- 2.04    &  3519.3     & 113.4    &  105.6 &  3819.2     & 123.5    &  103.4 \\
   2.04 -- 2.20    &  3668.8     & 130.3    &  118.0 &  3958.6     & 134.6    &  131.7 \\
   2.20 -- 2.38    &  3566.2     & 84.6    &  79.0 &  3799.8     & 120.1    &  119.8 \\
   2.38 -- 2.57    &  3605.2     & 77.7    &  75.7 &  3978.6     & 108.3    &  111.6 \\
   2.57 -- 2.78    &  3734.5     & 135.9    &  125.1 &  3874.1     & 80.7    &  86.5 \\
   2.78 -- 3.00    &  3742.5     & 143.3    &  123.6 &  3843.1     & 129.6    &  113.4 \\
   3.00 -- 3.24    &  3667.5     & 128.1    &  116.3 &  3883.0     & 99.1    &  99.2 \\
   3.24 -- 3.50    &  3529.4     & 92.7    &  81.3 &  3731.5     & 124.8    &  104.9 \\
   3.50 -- 3.79    &  3606.2     & 99.2    &  94.2 &  3918.2     & 103.2    &  108.1 \\
   3.79 -- 4.09    &  3564.3     & 99.5    &  91.9 &  3960.4     & 101.6    &  95.2 \\
   4.09 -- 4.41    &  3682.4     & 107.1    &  103.8 &  3845.0     & 82.7    &  109.2 \\
   4.41 -- 4.76    &  3750.4     & 68.2    &  68.0 &  3674.6     & 113.3    &  89.8 \\
   4.76 -- 5.20    &  3689.0     & 134.6    &  110.9 &  3562.4     & 75.6    &  75.2 \\
\hline
\hline
\end{tabular}
\caption{\label{tab:merged_transm_spec} NIRSpec/PRISM spectrum of TRAPPIST-1\,d for visits 1 and 2. These spectra are obtained directly from the transit light curves, without correction for the impact of stellar contamination. A machine-readable version of this table will be made available in the electronic journal.}
\end{table}

\setcounter{figure}{0}
\renewcommand{\thefigure}{C\arabic{figure}}
\setcounter{table}{0}
\renewcommand{\thetable}{C\arabic{table}}

\section{Out-of-transit and in-transit constraints on stellar heterogeneities}

We constrain the coverage of bright and dark heterogeneities on the visible stellar hemisphere from two independent methods: retrievals on each visit's out-of-transit stellar spectra, and on the planet's transmission spectra affected by unocculted stellar surface heterogeneities.

Throughout this study, we use the words ``facula(e)'' (``spot(s)'') to refer to regions of the stellar surface that are well represented by the 1D spectrum of a hotter (cooler) star than the photosphere, i.e., brighter or dimmer at all wavelengths (following e.g. \citealp{rackham_transit_2018,iyer_influence_2019}). However, we recognize that this terminology is limited since the nature of stellar surface heterogeneities on late-type M dwarf stars can be drastically different than their F, G, or even K dwarf counterparts. For example, modeling work (see e.g. \citealp{beeck_3d_2015,norris_2023}) has suggested that magnetic activity could result in the formation of \textit{dark} faculae with bright hot walls on the surfaces of M dwarfs in-between granules. Such 3D modeling of faculae contrast spectra also revealed that they can be both brighter and dimmer than the quiet photosphere depending on the wavelength range considered. This means that our simplified models leveraging 1D grids may lead to skewed inferences on the coverage fractions of different heterogeneities at the stellar surface, and we recognize that the interpretability of our results would benefit from further model fidelity. However, without any large grids of late-M heterogeneity contrast spectra available, as required to properly model these effects, we default to the current state-of-the-art for stellar surface heterogeneity considerations in transit spectroscopy as described in Sections \ref{sec:oot_spectrum_methods} and \ref{sec:TLS_methods}. 

\subsection{Flux-calibrated out-of-transit stellar spectrum}\label{sec:fluxcal} 

\begin{figure}
    \centering
    \includegraphics[width=0.85\textwidth]{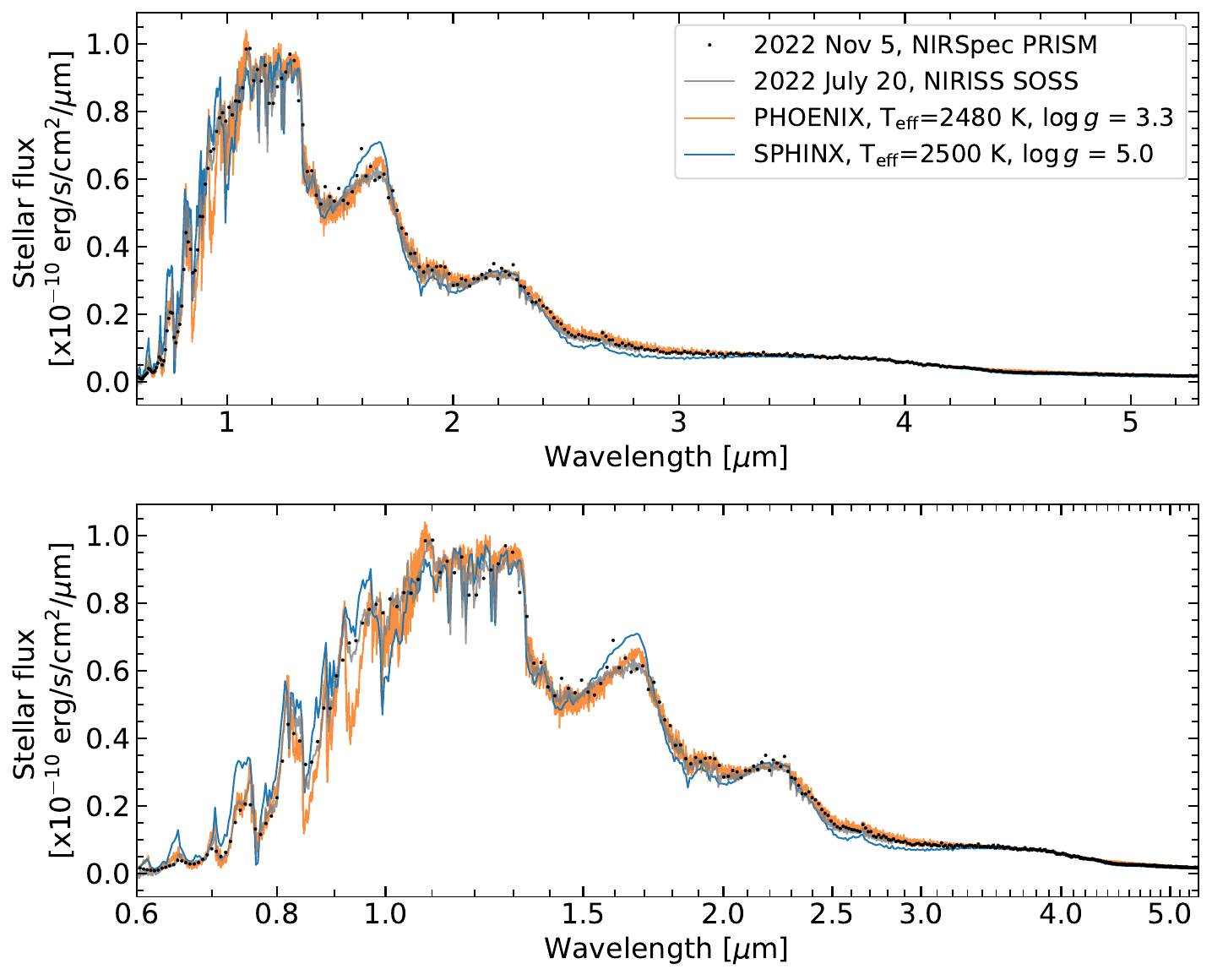}
    \vspace{5mm}
    \vspace{-5mm}\caption{Comparison of the NIRSpec/PRISM spectrum of TRAPPIST-1 (black points) with NIRISS/SOSS observations (gray line, \citealp{lim_atmospheric_2023}) and stellar models. We show the PHOENIX model in our grid with the $T_\mathrm{eff}$ and $\log g$ combination that provides the closest match to our NIRSpec/PRISM spectrum in orange (convolved to R=2,000), and a SPHINX model with $T_\mathrm{eff}=2500$\,K and $\log g=5$ \citep{iyer_sphinx_2022,sphinx_zenodo}. Both PHOENIX models and SPHINX models struggle to reproduce the short-wavelength part of the spectrum, and the SPHINX model additionally does not capture well the relative depths of the broadband spectral features observed past 1.5$\mu$m.}
    \label{fig:oot_phoenix_sphinx}
\end{figure}

For each visit, we perform a modified version of the data reduction in order to obtain the time-series of flux-calibrated stellar spectra. The steps are identical to the standard data reduction described in Section \ref{sec:eureka}, except for a few differences in Stages 2 and 3. Stage 2 is performed using \texttt{Eureka!}, except that the flat-field and absolute photometric calibration steps (\texttt{FlatFieldStep} and \texttt{PhotomStep} in the \texttt{jwst} pipeline) are not skipped. We also modify the Stage 3 \texttt{Eureka!} code where the optimal spectral extraction is performed in order to mask rows that are more than 8 pixels away from the trace, following \citet{moran_high_2023}.

We then select integrations from which to construct the out-of-transit spectra. Since our analysis aims at determining the combination of stellar photosphere and surface heterogeneity components that best describe the ``quiet'' photosphere, we discard not only in-transit integrations, but also out-of-transit integrations which coincide with likely flare events from the H$\alpha$ time series and the shape of the fitted GP component (Figure \ref{fig:wlc_Halpha}). Specifically, for visit 1, we only consider integrations prior to BJD$_\mathrm{UTC}=2459889.245$, and for visit 2, we discard integrations in BJD$_\mathrm{UTC} \in [2459893.245,2459893.27]$ and BJD$_\mathrm{UTC}>2459893.295$. We then compute the median in time of all the spectra from the remaining integrations to create the out-of-transit spectrum for each visit, and set as the error bar on each point the standard deviation in that wavelength band over all the integrations considered. Finally, when performing the out-of-transit retrieval on the flux-calibrated spectra, we also crop all wavelengths greater than 5.5$\mu$m which is the long-wavelength edge of the PHOENIX models \citep{husser_new_2013} that we use to construct the model spectra (see Section \ref{sec:stellar_ctm}). 

\subsection{Out-of-transit stellar spectrum modeling}\label{sec:oot_spectrum_methods}

\texttt{exotune} models the out-of-transit stellar spectrum as a linear combination of 1, 2, or 3 spectral components, each defined by its effective temperature $T_\mathrm{comp}$ and $\log g_\mathrm{comp}$. The 1-component model assumes that the entire spectrum is described by the quiet photosphere, while 2 or three-component models simulate the emergent spectrum as the linear combination of 2 or 3 stellar spectra, with one corresponding to the photosphere, and the other(s) describing a cooler ``spot'' component (covering a fraction $f_\mathrm{spot}$ of the stellar surface) and/or a hotter ``facula'' component (covering a fraction $f_\mathrm{fac}$ of the visible hemisphere on the stellar surface). Our model out-of-transit stellar flux $F_\mathrm{oot}$ is therefore constructed as:

\begin{eqnarray}
        F_\mathrm{oot} = \left(1-f_\mathrm{spot}-f_\mathrm{fac}\right) \times F_\mathrm{phot} \nonumber \\
    + f_\mathrm{spot}\times F_\mathrm{spot}+ f_\mathrm{fac}\times F_\mathrm{fac}
\end{eqnarray}

\noindent where $F_\mathrm{phot}$ ($T_\mathrm{phot}$,$\log g_\mathrm{phot}$),  $F_\mathrm{spot}$($T_\mathrm{spot}$,$\log g_\mathrm{spot}$), and $F_\mathrm{fac}$($T_\mathrm{fac}$,$\log g_\mathrm{fac}$) are stellar surface fluxes computed at R=10,000 from the PHOENIX library \citep{husser_new_2013} obtained with the \texttt{MSG} module \citep{townsend_msg_2023}. This forward model is implemented in \texttt{exotune} within a retrieval framework by pre-computing a grid of stellar models in increments of 10\,K in $T_\mathrm{eff}$ and 0.1 in $\log g$ and choosing at each iteration and for each component the closest model in the grid. Once the high-resolution models are combined to obtain $F_\mathrm{oot}$, we integrate the flux within each wavelength bin of the NIRSpec/PRISM spectrum to compute the Gaussian likelihood within a Bayesian framework. We adopt uniform priors on $f_\mathrm{spot}$ and $f_\mathrm{fac}$ requiring  $f_\mathrm{spot}+f_\mathrm{fac}<1$, on $T_\mathrm{phot} \in [2300,3700]$ (2300\,K being the lowest temperature in the PHOENIX model grids), and on the temperature contrasts of the hot and cool heterogeneity components, $|\Delta T_\mathrm{het}| \in [50,1000]$ ($\Delta T_\mathrm{fac}>0$, $\Delta T_\mathrm{spot}<0$, $T_\mathrm{het} = T_\mathrm{phot} + \Delta T_\mathrm{het}$). We also set a uniform priors on the photosphere $\log g \in [2.5,5.5]$ (since PHOENIX models with a value of $\log g \sim 3.5$, much lower than the literature value, were found to provide a better match to the NIRISS/SOSS transmission spectrum of TRAPPIST-1 from a forward model exploration; \citealp{lim_atmospheric_2023}; Figure \ref{fig:oot_phoenix_sphinx}). We similarly adopt a uniform prior on the difference $\Delta \log g_\mathrm{het}<0$ between the heterogeneity and the photosphere component; such a difference could be induced by magnetic activity (see e.g. \citealp{fournier-tondreau_near-infrared_2023}). Additionally, we fit a factor to scale $F_\mathrm{oot}$ to the flux of TRAPPIST-1 received at Earth, and an error inflation factor $f_\mathrm{errInfl} \in [0,50]$ that multiplies uniformly the estimated errors on the measured flux to account for a potentially imperfect model fidelity. 

We explore the parameter space using the MCMC sampler \texttt{emcee} \citep{foreman-mackey_emcee_2013}, within a fully parallelized framework. We use 20 times as many walkers as there are fitted parameters. Each chain is run for 100,000 steps, 60\% of which are ignored (burn-in) to obtain the final posterior distributions on the heterogeneity parameters. We ensure that the chains run for more than 50 times the largest autocorrelation timescale inferred across all parameters ($<840$ steps across all our runs). Motivated by the downward slopes in the transmission spectra at the blue end of the NIRSpec/PRISM spectrum (Fig. \ref{fig:fitted_stctm_spectra}), we perform 2- (photosphere and faculae) and three-component (photosphere, spots and faculae) retrievals on the out-of-transit stellar spectrum for each visit. We record the Bayesian Information Criterion (BIC) value for each retrieval for model comparison purposes.

\subsection{Transit light source effect modeling}\label{sec:TLS_methods}

We further use \texttt{stctm} (\citealp{stctm_zenodo_temp}, Piaulet-Ghorayeb, subm.) to model the transmission spectrum of TRAPPIST-1\,d, assuming that the planet is either a bare rock or that its transmission spectrum is otherwise featureless. We model the transit light source effect assuming two populations of heterogeneities referred to as spots and faculae, and parameterized as previously described in \citet{Piaulet_Ghorayeb_2024} and in the previous section. We also perform retrievals where only one population of heterogeneities (faculae) is considered, in order to assess the impact of degeneracies between spot and facula heterogeneity properties on the final posterior distributions we obtain. 

In our retrievals, we fit the temperatures of the heterogeneity components following the same parameterization as in the \texttt{exotune} fit to the out-of-transit stellar spectrum. We similarly use a pre-computed fine grid of stellar models to compute the forward models for each model evaluation.

The parameter space is sampled using \texttt{emcee} \citep{foreman-mackey_emcee_2013}, using 20 times as many walkers as there are fitted parameters, and running the chains for 100,000 steps, with 60\% discarded as burn-in, confirming that each chain ran for more than 50 times the maximum estimated autocorrelation timescale (ranging from 500 to 1700 steps depending on the run). For the heterogeneity fractions, we test both uniform and log-uniform priors, always ensuring that $f_\mathrm{spot}+f_\mathrm{fac}<1$ in order to assess whether the data could provide a lower bound on $f_\mathrm{fac}$ given the strong downwards slope observed in the visit 2 spectrum (Fig. \ref{fig:fitted_stctm_spectra}). The model spectra and stellar heterogeneity parameters are consistent regardless of the prior we adopt, and we report the results from the fit performed with a log-uniform prior on the facula covering fraction for the 2- and three-component retrievals (since we recover a lower limit for visit 2), and with a linear prior on the spot covering fraction (which is unconstrained regardless of the prior we adopt) for the three-component fit.
We adopt uniform priors on the heterogeneity temperature contrasts $\Delta T_\mathrm{het}$, with $|\Delta T_\mathrm{het}|>100$ to ensure that the heterogeneity spectra do not default to the photosphere spectrum. Contrary to the out-of-transit stellar spectrum fit, the TLS retrievals are not sensitive to the photosphere temperature, but only to the \textit{ratios} between the photosphere and heterogeneity component spectra. Therefore, we impose a Gaussian prior on $T_\mathrm{phot}$ using the median effective temperature value reported by \citet{agol_refining_2021} and a wider standard deviation of 70\,K, following the procedure used for the \texttt{stctm} fit to the transmission spectra of TRAPPIST-1\,b \citep{lim_atmospheric_2023} and c \citep{RadicaT1c}. We implement a new functionality in \texttt{stctm} that allows us to also fit the $\log g$ of the photosphere (same prior as the \texttt{exotune} retrieval) and the $\Delta \log g$ contrast between the surface gravity used to describe the heterogeneity and photosphere components, in a way analogous to \texttt{exotune}. Finally, we fit the wavelength-independent transit depth of the planet itself, $D_\mathrm{scale, TLS}$, with a wide uniform prior. We record the BIC value in each case in order to evaluate to what extent higher-complexity models are favored by the data.

\begin{table}[h!]
\centering
\caption{Results from the three-component retrievals performed with \texttt{stctm} on the (in-transit) transmission spectra, and on the out-of-transit stellar spectra. We report $1\sigma$ confidence intervals for each parameter, or $2\sigma$ upper or lower limits where applicable. We use the posterior samples of $\Delta T_\mathrm{spot}$, $\Delta T_\mathrm{fac}$, and $ T_\mathrm{phot}$ to obtain the values quoted for $T_\mathrm{spot}$ and $T_\mathrm{fac}$.}
\label{tab:retrieved_stctm_params}
\begin{threeparttable}
\begin{tabular}{c|cc|cc}
    \hline
    \hline
    \multirow{2}{*}{Parameter} & \multicolumn{2}{c}{Visit 1} & \multicolumn{2}{c}{Visit 2} \\
              & In-transit (TLS) & Out-of-transit & In-transit (TLS) & Out-of-transit\\
    \hline
    $T_\mathrm{phot}$ [K]            & $2604.0^{+67.0}_{-64.0}$ & $2630.0^{+93.0}_{-69.0}$ & $2557.0^{+60.0}_{-54.0}$ & $2639.0^{+111.0}_{-72.0}$ \\
    $\log g_\mathrm{phot}$           & $3.4^{+0.9}_{-0.4}$ & $4.0^{+0.6}_{-0.4}$ & $4.8^{+0.3}_{-0.5}$ & $4.0^{+0.7}_{-0.5}$  \\
    $T_\mathrm{spot}$ [K]            & $2388.0^{+90.0}_{-60.0}$ & $2418.0^{+62.0}_{-73.0}$ & $2357.0^{+58.0}_{-40.0}$ & $2415.0^{+59.0}_{-67.0}$  \\
    $f_\mathrm{spot}$                & $0.18^{+0.25}_{-0.14}$ & $0.64^{+0.17}_{-0.32}$ & $0.40^{+0.18}_{-0.18}$ & $0.68^{+0.15}_{-0.32}$ \\
    $T_\mathrm{fac}$ [K]            & $3281.0^{+163.0}_{-248.0}$ & $3127.0^{+268.0}_{-272.0}$ & $2957.0^{+215.0}_{-150.0}$ & $3077.0^{+273.0}_{-249.0}$ \\
    $f_\mathrm{fac}$  & $0.007-0.076$ & $<0.10$ & $0.058-0.274$ & $<0.089$ \\
    $\Delta \log g_\mathrm{het}$       & $>-1.3$ & $>-2.1$ & $>-0.6$ & $>-2.2$  \\
    \hline
    \hline
\end{tabular}
\end{threeparttable}
\end{table}

\subsection{Constraints from the transit light source effect}\label{sec:TLS_results}

Both spectra exhibit telltale signs of stellar contamination (Figure \ref{fig:fitted_stctm_spectra}), in particular the downward slope shortwards of about 1$\mu$m which can only be reproduced by the dilution effect of unocculted bright inhomogeneities on the stellar surface. The good match to the spectra with the \texttt{stctm} retrievals support this, as we can reproduce most of the features without the need for a planetary atmosphere (Figure \ref{fig:fitted_stctm_spectra}). The 4-5$\mu$m slope in the visit 2 transmission spectrum could not be accurately matched by our models, and would also be surprising from an atmospheric standpoint. We do not find motivation in the spectroscopic light curves to discard the two redmost points. Still, only 1 point cannot be reproduced within 1 sigma by the \texttt{stctm} model, and we confirm that running the SCARLET atmosphere retrievals with or without that point does not substantially impact our inferences.

Overall, the Visit 2 spectrum is the most affected by stellar contamination, with the measured transit depth at 0.7$\mu$m being about 1000 ppm lower than that measured at 1.3$\mu$m. Beyond the short-wavelength slope, the median transit depth from 1.2 to 5$\mu$m is also higher in the second visit compared to the first while we infer consistent orbital parameters across visits (Figure \ref{fig:b_aRs}), supporting the interpretation of visit-to-visit variations in stellar contamination (Figure \ref{fig:fitted_stctm_spectra}). 


The short-wavelength slopes in the spectra favor the inclusion of faculae rather than assuming only a single-temperature photosphere (1.8$\sigma$ for the visit 1 TLS fit, 2.8$\sigma$ for visit 2 retrieval), but the contribution of spots to the fit, especially for visit 2 where large $f_\mathrm{spot}$ are favored, is non-trivial in three-component fits. We find that the $\Delta BIC$ disfavors adding the spot component for the first visit (the additional parameters are not warranted by the marginal improvement in fit quality). Meanwhile, the addition of spots to the visit 2 retrieval is only weakly supported with a $\Delta BIC$ of 2.3. Comparing the best-fit model from the three-component fits to a two-component (photosphere and faculae) alternative sheds some light on these results (Figure \ref{fig:fitted_stctm_spectra}, top panel). Generally, for both visits, the 2- and three-component models are identical shortwards of about 1.3$\mu$m (i.e., the downward slope is purely described by the facula component). However, the transit depth variations longwards of 1.3$\mu$m are better matched with the addition of the cool heterogeneity. For visit 1, spots provide a better match to the (tentative, $<1\sigma$) $1.5$ to $2\mu$m dip and to the 2.5 to 3$\mu$m bump in the spectrum, while the model accounting only for faculae produces a flat spectrum in these regions. Conversely, for visit 2, the faculae-only model predicts, given the large $f_\mathrm{fac}$, that one would observe TLS-imprinted ``inverse'' water features peaking near 1.4, 1.8, and 2.8$\mu$m, while the observed spectrum is much flatter, hence the role played by the spot component.

\subsection{Characteristics of the out-of-transit stellar spectrum}\label{sec:oot_results}

The \texttt{exotune} out-of-transit stellar spectrum retrievals to each individual visit yield consistent conclusions. Both visits can be explained by a $\sim 2500$\,K photosphere, a minor faculae component with 2$\sigma$ upper limits on $f_\mathrm{fac}$ of about 10\%, and a potential spot component. Since two-component (photosphere and faculae) retrievals provided a similarly good match to the data, while inferring lower photospheric temperatures closer to 2450\,K, we hypothesize that in the case of the three-component fit, the preferred large spot covering fraction essentially buffers for the fact that the bulk of the spectrum could be explained by a cool, 2450-2500\,K component. For either visit, the $\Delta BIC$ does not significantly favor the inclusion of the spot component in addition to the faculae.

These conclusions hold whether or not we include in \texttt{exotune} fit the columns affected by partial saturation. Over these wavelengths, our stellar spectrum is also consistent with the one reported with NIRISS/SOSS \citep{lim_atmospheric_2023}, which was not affected by saturation (see Figure \ref{fig:oot_phoenix_sphinx}). 

\subsection{Potential caveats of the stellar surface modeling}\label{sec:caveats_oot_TLS}

The reliability of stellar surface inferences for TRAPPIST-1 from stellar models using the PHOENIX model grid deserves further scrutiny. First, stellar model fidelity for late M dwarfs is imperfect, and may call for data-driven approaches given important mismatches between observations and models at the level of stellar SEDs, as well as missing lines in the reference line lists \citep{Rackham_SAG21_2023,deWit_roadmap_2023, davoudi_updated_2024,jahandar_chemical_2024}.  

We find, consistent with previous work \citep{lim_atmospheric_2023}, that we need to use the stellar photosphere $\log g$ as a fudge factor to accurately match the stellar spectrum with PHOENIX models (Figure \ref{fig:oot_phoenix_sphinx}). Both TLS and stellar spectrum retrievals favor low $\log g \sim 3.5$-4.0 (Table \ref{tab:retrieved_stctm_params}) that are inconsistent with the literature inferences for TRAPPIST-1, but allow us to better reproduce the observed spectral features. We note that, although low $\log g$ values are favored by our analysis, the low resolution of NIRSpec/PRISM relative to NIRISS/SOSS reduces our sensitivity to the two main features near 1.05 and 1.6$\mu$m that drive the inference on $\log g$ (Figure \ref{fig:oot_phoenix_sphinx}). Future studies aiming at characterizing the star should ideally use both NIRISS and NIRSpec in conjunction, in order to achieve high enough resolving power at the short wavelengths while capturing the long-wavelength Rayleigh-Jeans tail of the spectrum.

As an alternative to PHOENIX models, we considered the SPHINX \citep{iyer_sphinx_2022,sphinx_zenodo} model grid. We compare our spectrum to a nominal SPHINX model for $T_\mathrm{eff}=2500$\,K,  $\log g=5.0$, solar $[\mathrm{Fe/H}]$ and C/O ratio, and find that although some spectral regions near 1.1 to 1.3 microns are slightly better fit by this model, others at shorter wavelengths show a similar mismatch to the data as PHOENIX models do, and further fail to reproduce the relative strengths of the broadband spectral features observed past 1.3$\mu$m. We do not perform retrievals using SPHINX models, since their $R\sim 250$ resolving power is too low to be further leveraged by joint TLS-atmospheric retrievals, which reduces their interest in terms of contextualizing findings about the extent to which stellar contamination affects the transmission spectrum. 

We also recognize as a potential caveat the fact that the stellar effective temperature, near 2600\,K, is only 300\,K away from the lowest temperature covered by the PHOENIX model grids, at 2300\,K. More detailed models tailored to the conditions on low-mass M dwarfs could therefore help tease out more stellar surface information content from future observations.

\setcounter{figure}{0}
\renewcommand{\thefigure}{D\arabic{figure}}
\setcounter{table}{0}
\renewcommand{\thetable}{D\arabic{table}}

\section{Details of the atmosphere modeling methodology}\label{sec:atm_model_methods}

\subsection{Sequential retrievals of atmospheric properties}\label{sec:sequential_SCARLET}

We use the SCARLET \citep{benneke_distinguishing_2012,benneke_characterizing_2013,benneke_strict_2015,benneke_sub-neptune_2019, benneke_water_2019, pelletier_where_2021, piaulet_evidence_2023} 1D atmospheric forward modeling and retrieval framework to interpret the observed transmission spectrum of TRAPPIST-1\,d. With SCARLET, we fit the combined (visit 1 + visit 2) stellar-contamination-corrected transmission spectrum, as different stellar contamination parameters would be required to describe each visit for a joint fit.
We perform free chemistry retrievals tailored to probe thin, potentially high mean molecular weight atmospheres of rocky planets which we had previously applied to TRAPPIST-1\,c \citep{RadicaT1c}, and TRAPPIST-1\,g (Benneke et al., submitted). 

In SCARLET, the forward model iterates over the radiative transfer and hydrostatic equilibrium calculations for each sampled atmospheric composition and temperature structure to find the best-matching planetary radius at 10 mbar. Overall, our suite of retrievals included H$_2$, He, N$_2$, CH$_4$, H$_2$O, CO, CO$_2$, NH$_3$, and SO$_2$. For absorbing species, we use the HELIOS-K computed cross-sections of H$_2$O \citep{ExoMol_H2O}, 
 CO \citep{Hargreaves2019}, CO$_2$ \citep{ExoMol_CO2},  CH$_4$ \citep{HargreavesEtal2020apjsHitempCH4}, NH$_3$ \citep{ExoMol_NH3}, and SO$_2$ \citep{ExoMol_SO2}. Our forward models are computed at R=31,250, and binned to the resolution of the data for the likelihood calculation within our Bayesian inference framework. 

We explored a range of parameterizations for the atmospheric composition to explain the spectrum of TRAPPIST-1\,d. For one set of models, we consider single-molecule atmospheres (100\% CH$_4$, H$_2$O, CO$_2$, CO, or NH$_3$) where we fit for the effective surface pressure $p_\mathrm{surf}$ (which could be the bare rock surface or top of an opaque cloud deck). We also test scenarios where N$_2$ is combined with one of CH$_4$ or CO$_2$ (fitting for the partial pressure of each component), as seen in the atmospheres of some solar system terrestrials. Finally, we also calculate models where all molecules (CH$_4$, CO$_2$, CO, NH$_3$, H$_2$O, SO$_2$) are together in a multi-gas atmosphere with H$_2$ as the background gas. We do not consider more complex models including aerosols with variable size distribution, as we do not detect significant wavelength-dependent trends in our data. For the temperature profile, we fit for the temperature of the photosphere at the terminator, which is representative of the region probed by our transmission observations. We explore the parameter space using Nested Sampling, as implemented in the Python module \texttt{nestle}  \citep{skilling_nested_2004, skilling_nested_2006} and use at least 3,000 live points per retrieval in the final results we present.

We further calculate a set of seven representative forward models using SCARLET for visual comparison to our observed spectrum: six corresponding to cloud-free 1-bar atmospheres fully composed of one of the absorbers we tested for (CH$_4$, CO, H$_2$O, NH$_3$, CO$_2$, SO$_2$) and one representative of a cloud-free, solar-composition atmosphere in chemical equilibrium (composition from \citealp{stock_fastchem:_2018}).

\subsection{Joint retrievals of stellar contamination and planetary atmosphere contribution} \label{sec:joint_star_planet_retrieval}

The atmosphere-only retrievals presented above provide a first set of ``sequential'' retrievals, where the planetary atmosphere contribution is constrained from retrievals performed on the transmission spectrum of TRAPPIST-1\,d after it has been corrected for the stellar contamination contribution constrained by \texttt{stctm}. We also perform more computationally expensive joint retrievals where we fit simultaneously for the contribution of unocculted stellar heterogeneities, and of a potential planetary atmosphere, to the observed spectrum of each visit. This second scheme is more conservative, as it marginalizes over the interplay between stellar and planetary contributions to the transmission spectrum instead of only using one realization of the stellar contamination signal. Further, planetary and stellar features could effectively ``cancel out'' to form a flat spectrum, which means that the data could allow for a broader range of atmosphere scenarios than afforded by the stellar-contamination-corrected spectrum. 

In these joint retrievals, for each visit we fit the covering fractions of the cool and hot heterogeneities, and the differences between the photosphere temperature and that of each of the two heterogeneity populations. Informed by the TLS-only and out-of-transit spectral retrieval results, we fit the stellar $\log g$ but do not fit any $\log g$ difference between the photosphere and heterogeneity component. We also fit for an offset to account for a potential difference in the broadband transit depth between the two visits that could be introduced from left-over flux from the in-transit flare in visit 1 after the transit, affecting the baseline and acting as a dilution factor (as seen in \citealp{RadicaT1c}). This results in a total of eleven additional fitted parameters. 

We use uniform and uniformative priors on each parameter, except for the photosphere temperature where we use the same Gaussian prior as in the \texttt{stctm} fit.







\subsection{Impact of refraction on sensitivity to planetary atmospheres} \label{sec:refraction}

Atmospheric refraction occurs when starlight passes through an exoplanet's atmosphere and bends due to gradients in the refractive index. In transmission spectroscopy, this bending of light can impose a refraction limit, effectively setting a boundary below which stellar light is refracted away from the observer's line of sight \citep{misra_effects_2014,robinson_analytic_2017,betremieux_hidden_2018}. As a result, only the higher, less dense layers of the atmosphere contribute to the transmission spectrum. This phenomenon can restrict our ability to probe the deeper atmospheric regions. and is therefore crucial to consider when interpreting transmission spectra. 

We use the Planetary Spectrum Generator (PSG, \citealp{Villanueva2018, Villanueva2022}) to compute the refraction limit from 3D GCM simulations of TRAPPIST-1\,d's atmospheric conditions \citep{turbet_water_2023}. PSG computes the refraction accounting for a non-uniform spherical atmosphere through numerical ray tracing. The light is traced from the observer through the planet's atmosphere to the surface, or the top of the atmosphere, modulated by some adjustments of the incidence angles at each layer interface as a function of the refractive properties of the layers. The refraction coefficient is influenced by the layer composition, local density and wavelength of the radiation. Since our spectrum rules out thick atmospheres for hydrogen-dominated atmospheres (Figure \ref{fig:met_psurf}), we focus on potential secondary atmosphere compositions rich in volatiles, where a broader range of surface pressures can be invoked. We select two edge-case atmospheric scenarios in terms of the atmospheric mean molecular weight since it correlates with the refractive index (higher for heavier species): a water-dominated (10 bar H$_2$O, 1 bar N$_2$) composition, and a pure-CO$_2$ (10 bar CO$_2$) composition. The surface pressures in these models lie at the upper bound of our priors for the retrievals that explore high mean molecular weight atmospheres specifically (Figures \ref{fig:atmosphere_molbymol} and \ref{fig:partialps}). This test allows us to assess whether the limits we place across all the atmospheric scenarios we explore indeed lie within the range of pressures we would be sensitive to.

\subsection{High mean molecular weight atmosphere retrievals}

We report the constraints obtained from single-molecule retrievals in Table \ref{tab:atm_table}. 
\begin{table*}
\centering
\begin{tabular}{lccc}
\hline
\hline
Atmospheric scenario & Sequential fit & Joint fit \\
\hline

Composition & \multicolumn{2}{c}{$\log_{10} p_\mathrm{surf,eff}$ [bar]} \\

\hline
100\% CH$_4$ & $<-4.66$ & $-4.77$\\
100\% CO & $<-2.55$ & $-3.24$ \\
100\% H$_2$O & $<-2.42$ & $<-2.06$ \\
100\% NH$_3$ & $<-3.09$ & N/A \\
100\% CO$_2$ & $<0.52$ & $<-1.44$ \\
100\% SO$_2$ & $<0.32$ & $<-1.95$ \\
\hline
\end{tabular}
\caption{\label{tab:atm_table}Constraints on the atmospheric composition of TRAPPIST-1\,d from the SCARLET retrievals for cases with high atmospheric mean molecular weight.
For each of the the single-molecule retrievals}, we report the upper limit at 95\% confidence obtained on the effective surface pressure in either the sequential fit or the joint stellar contamination+planetary atmosphere retrieval setup. 
\end{table*}




\setcounter{figure}{0}
\renewcommand{\thefigure}{E\arabic{figure}}
\setcounter{table}{0}
\renewcommand{\thetable}{E\arabic{table}}

\section{Water loss simulations for TRAPPIST-1\,d, e, f, and g}\label{sec:water_loss_methods}

We use the box model described in \citet{moore_role_2023}, including the planetary-mass-dependence outlined in \citet{moore_water_2024}, to assess the vulnerability of TRAPPIST-1\,d to desiccation as a function of its initial water inventory. The model assumes that the planet remains at the same orbital distance throughout its evolution. The planet experiences time-variable water loss modulated by the amount of stellar irradiation it receives and by its evolving internal state which dictates the amount of surface water susceptible to escape.

The stellar XUV irradiation is assumed to follow evolutionary models for low-mass stars \citep{baraffe_new_2015}. We model the XUV luminosity as initially saturated for 1 Gyr \citep{ribas_evolution_2005}, and account for a 5 Myr delay between star and planet formation, offsetting the stellar tracks accordingly to account for the cumulative XUV flux received by the planetary atmosphere. 

The initial water inventory of the planet begins completely dissolved within the magma ocean. As the magma ocean solidifies, water is partitioned between three reservoirs: the magma ocean, the solidified mantle, and the planetary surface (where, due to runaway greenhouse surface temperatures, water exists in a steam atmosphere). Due to the high solubility of water in the initially fully molten mantle, a large fraction of the water is partitioned into the melt until the ``bottom-up'' solidification of the mantle is complete (see e.g., \citealp{elkins-tanton_ranges_2008,dorn_hidden_2021}). The complete solidification of the mantle coincides with the end of the runaway greenhouse phase, which is set by the level of irradiation received by the planet. Once the mantle is solidified, we model deep water cycling --- where surface temperatures are now low enough that water can exist in condensed form on the surface --- similar to that experienced on Earth due to plate tectonics \citep{mcgovern_1989} over geological timescales. We note that the magma ocean could instead solidify before the end of the runaway gas accretion phase (see \citealp{selsis_cool_2023}), but our simple model does not account for non-adiabatic thermal profiles that result in this early mantle solidification.

During the runaway greenhouse phase, water loss to space is modeled following energy-limited escape \citep{watson_dynamics_1981,erkaev_roche_2007,lopez_born_2017}. The escape regime transitions to diffusion-limited escape \citep{hunten_escape_1973} once surface water condensation becomes possible -- with hydrogen slowly diffusing to the exobase where it can escape to space. 

We model the evolution of a planet with the mass of TRAPPIST-1\,d (Table \ref{table:pla_star_param}) at several fixed orbital distances ranging from 0.2229 AU to 0.0526 AU (to account for the uncertainty on the formation location and migration timescale of the planet) and with varying initial amounts of volatiles ranging from about 0.75 to 8 Earth oceans, which is reasonable for TRAPPIST-1\,d \citep{krissansen-totton_predictions_2022}. We model the water partitioning, water loss and interior evolution of the planet over 5.4 to 9.8 Gyr, to account for the uncertainty on the stellar age \citep{burgasser_age_2017}. Additionally, we investigate the water loss outcomes for TRAPPIST-1 e, f, and g, assuming initial volatile inventories ranging from 0.5 to 12 Earth oceans and assuming they remained at their present orbital distance for 5.4 to 9.8 Gyr of evolution.

\begin{figure*}
    \centering
    \includegraphics[width=0.6\linewidth]{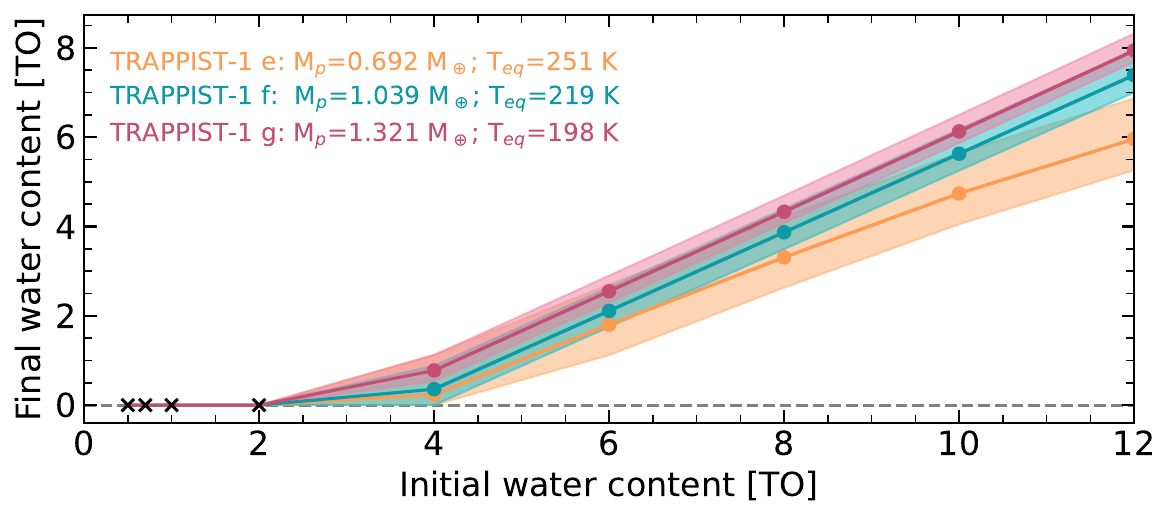}
    \vspace{5mm}
    \vspace{-5mm}\caption{Water inventory evolution simulation outcomes for TRAPPIST-1 e, f, and g. The final global water inventories are shown for TRAPPIST-1\,e (orange), f (blue) and g (pink), for the median stellar age of 7.6 Gyr (solid lines) and for the envelope of solutions over the range of stellar ages from 5.4 to 9.8 Gyr (semi-transparent shaded regions). The horizontal dashed line represents the limit of desiccation, where a simulation for a planet that ends up completely devoid of water would lie (black crosses). For planets e, f, and g, in our simulations, water inventories of more than $\sim 4$ Earth oceans (TO) are enough for the planet to avoid complete desiccation to this day. The quoted planet mass equilibrium temperature values are taken from \citet{agol_refining_2021}.}
    \label{fig:water_loss_efg}
\end{figure*}

The results for TRAPPIST-1\,d are shown in Figure \ref{fig:water_retention}. We present the results of our simulations in terms of the final (surface + atmosphere) water inventories on TRAPPIST-1 e, f, and g assuming in-situ evolution, for various initial volatile endowments (Figure \ref{fig:water_loss_efg}).

\bibliography{references}{}
\bibliographystyle{aasjournal}



\end{document}